\documentclass[manuscript]{acmart} 
\AtBeginDocument{%
  \providecommand\BibTeX{{%
    \normalfont B\kern-0.5em{\scshape i\kern-0.25em b}\kern-0.8em\TeX}}}

\copyrightyear{2023}
\acmYear{2023}
\setcopyright{rightsretained}
\acmConference[IUI '23]{28th International Conference on Intelligent User Interfaces}{March 27--31, 2023}{Sydney, NSW, Australia}
\acmBooktitle{28th International Conference on Intelligent User Interfaces (IUI '23), March 27--31, 2023, Sydney, NSW, Australia}
\acmDOI{10.1145/3581641.3584076}
\acmISBN{979-8-4007-0106-1/23/03}

\usepackage{algorithm2e}
\SetKwComment{Comment}{// }{}
\SetKwInOut{Input}{Input}
\SetKwInOut{Output}{Output}
\SetKwFunction{Singletons}{Singletons}
\SetKwFunction{List}{List}
\SetKwFunction{ShortestPath}{ShortestPath}
\SetKwFunction{Normalize}{Normalize}
\SetKwFunction{GraphStories}{GraphStories}
\RestyleAlgo{boxed}
\mathchardef\mhyphen="2D


\begin{document}

\title{Mixed Multi-Model Semantic Interaction for Graph-based Narrative Visualizations}

\author{Brian Keith Norambuena}
\email{briankeithn@vt.edu}
\affiliation{%
  \institution{Virginia Tech}
  \city{Blacksburg}
  \state{Virginia}
  \country{USA}
}
 \additionalaffiliation{%
   \institution{Universidad Católica del Norte}
   \department{Department of Computing \& Systems Engineering}
   \city{Antofagasta}
   \country{Chile}
   \postcode{1270709}
 }

\author{Tanushree Mitra}
\email{tmitra@uw.edu}
\affiliation{%
  \institution{University of Washington}
  \city{Seattle}
  \state{Washington}
  \country{USA}
}
\author{Chris North}
\email{north@vt.edu}
\affiliation{%
  \institution{Virginia Tech}
  \city{Blacksburg}
  \state{Virginia}
  \country{USA}
}

\renewcommand{\shortauthors}{Keith Norambuena et al.}

\begin{abstract}
Narrative sensemaking is an essential part of understanding sequential data. Narrative maps are a visual representation model that can assist analysts to understand narratives. In this work, we present a semantic interaction (SI) framework for narrative maps that can support analysts through their sensemaking process. In contrast to traditional SI systems which rely on dimensionality reduction and work on a projection space, our approach has an additional abstraction layer---the structure space---that builds upon the projection space and encodes the narrative in a discrete structure. This extra layer introduces additional challenges that must be addressed when integrating SI with the narrative extraction pipeline. We address these challenges by presenting the general concept of Mixed Multi-Model Semantic Interaction (3MSI)---an SI pipeline, where the highest-level model corresponds to an abstract discrete structure and the lower-level models are continuous. To evaluate the performance of our 3MSI models for narrative maps, we present a quantitative simulation-based evaluation and a qualitative evaluation with case studies and expert feedback. We find that our SI system can model the analysts' intent and support incremental formalism for narrative maps.
\end{abstract}

\maketitle

\section{Introduction}
Narratives are systems of stories \cite{halverson2011master}---sequences of events connected in a coherent manner. Narratives are fundamental to the sensemaking process and our understanding of the world \cite{keith2022design}, as humans use them as a natural way to capture relationships between sequences of events, alongside the goals, motivations, and plans of actors \cite{finlayson2013military}. Narrative sensemaking tasks range from intelligence analysis \cite{endert2014human}, where analysts try to find hidden or implicit connections between events, to journalistic analysis of news narratives, where analysts seek to understand the big picture across many articles \cite{bradel2015big}. 

To help analysts in sensemaking tasks, scholars have developed visual analytics tools, which aid analysts in processing and understanding greater quantities of data \cite{cook2005illuminating}. These tools usually focus on different parts of the sensemaking process \cite{pirolli2005sensemaking}. For example, some tools focus on the synthesis loop \cite{wright2006sandbox} to help analysts generate hypotheses. Others focus on the foraging loop \cite{kang2009evaluating}, where the goal is to gather and select appropriate data for further analysis. 

In this work, we focus on the synthesis loop of the sensemaking process. In particular, we deal with extracting narratives from large sets of documents describing events (e.g., news articles) in the form of a graph structure---a \textit{narrative map}---that represents the system of storylines \cite{keith2020maps}. These composite structures can then be used to aid analysts in narrative sensemaking tasks by providing a higher-level view and explicit connections based on chronology \cite{keith2022design} that traditional visualization systems, such as those based on dimensionality reduction (DR) and clustering of documents, would not well capture.

Recent work has sought to develop computational models to assist in the narrative sensemaking process \cite{keith2022design}. However, current approaches are static and lack refinement based on user- or task-specific goals beyond basic interactions such as searching or emphasizing specific keywords \cite{keith2020maps,ansah2019graph,cai2019temporal,liu2017growing,tannier2013building,shahaf2010connecting}.
More specifically, we design and evaluate an interactive narrative sensemaking tool that integrates previous work on narrative extraction and representation \cite{keith2020maps, keith2021vis, keith2022design} with the semantic interaction (SI) framework---a framework for visual analytics that enables steering models by inferring data characteristics that are of interest to the user based on their interactions \cite{endert2012semantic, hodas2016adding, self2018observation}. Our tool's purpose is to help analysts in narrative sensemaking tasks by building a better narrative model through \textit{incremental formalism} \cite{shipman1999formalism}---the ability to learn incrementally over multiple iterations to produce better models. Our main hypothesis is that the proposed SI models effectively support incremental formalism for narrative map models.

In particular, recent developments \cite{endert2014human} have shown the capabilities of semantic interaction techniques in aiding the sensemaking process \cite{bian2020deepva} through human-AI collaboration \cite{wenskovitch2020interactive}. Thus, we seek to integrate semantic interaction methods with narrative maps in order to create an interactive AI narrative sensemaking framework that is capable of learning from the analyst. These techniques could prove useful to an expert user who needs to extract and understand evolving narratives, such as intelligence analysts trying to uncover the underlying connections between documents or journalists attempting to understand rapidly evolving news narratives \cite{keith2022design}. 

To date, no previous research has sought to integrate semantic interaction with narrative extraction and visualization. Thus, our research provides a first step towards creating a human-AI interaction system that aids in the narrative sensemaking process. Furthermore, our overarching research goal is to explore \textit{how to design an SI model for narratives}. This requires dealing with the issue of how to modify the machine learning pipeline of narrative extraction to support SI (i.e., how do we model user intent in our narrative extraction model?).

\begin{figure*}[!htb]
    \centering
    \includegraphics[width=\textwidth]{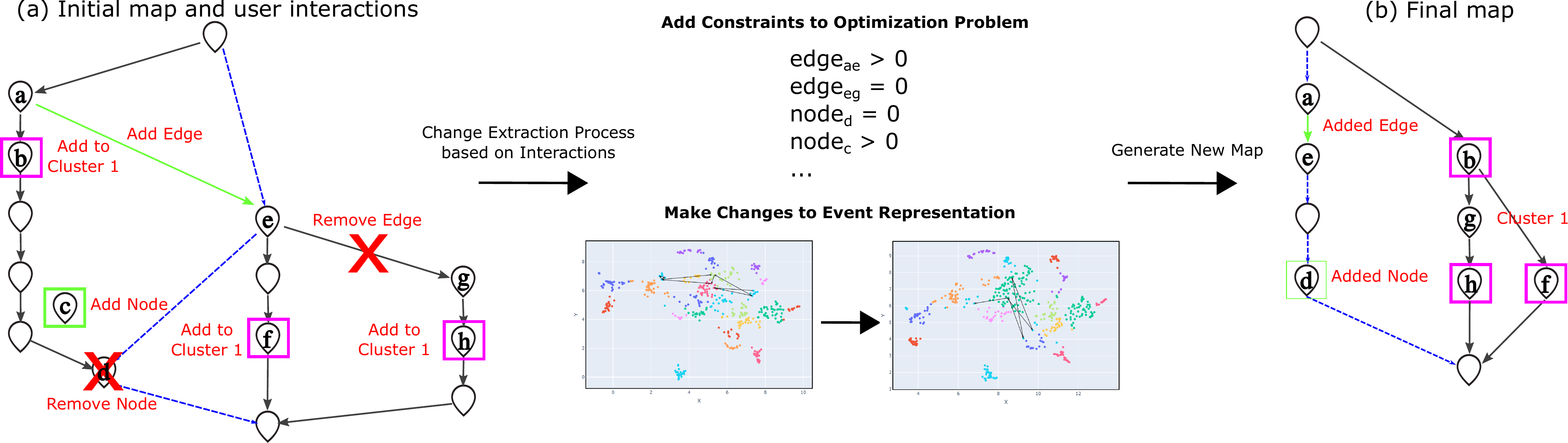}
    \caption{Overview of the semantic interaction model for narrative maps and its core interactions. The analyst modifies the original map on the left by: (1) adding an edge between events \textit{a} and \textit{e}, (2) adding event \textit{c}, (3) removing event \textit{d}, (4) removing the edge between events \textit{e} and \textit{g}, (5) grouping events \textit{b}, \textit{f}, and \textit{h} by assigning them to the same cluster. After this, the system makes changes to the extraction process based on the interactions and generates a new map that integrates the analyst's interactions into the model.}
    \label{fig:interactions}
\end{figure*}

As a specific example of how semantic interaction techniques could aid in the narrative sensemaking process, consider an analyst seeking to understand the causes and effects of a newsworthy event (e.g., the COVID-19 pandemic or social movements) from news data. The analysts would generate a visual representation of the narrative from data and provide their feedback through interactions (see Figure \ref{fig:interactions}), leading to incremental improvements as they work through their sensemaking process (e.g., removing biases, changing the focus of the narrative, or including specific storylines). We show a concrete example in Figure \ref{fig:covid} as part of a case study.

Furthermore, we note that traditional SI systems usually work directly on projection spaces generated by DR methods \cite{bradel2014multi}, leveraging distance changes induced by the user on the elements of the projection space to improve the underlying model. However, our proposed SI model has to deal with an additional challenge in the form of an intermediate abstraction layer: the structure space (i.e., the space of possible narrative graphs). Unlike the projection space, distance is not meaningful in the structure space. Instead, relationships are determined by the edges of the narrative graph. Thus, instead of continuous distance-based interactions, the natural interactions that arise in the structure space are discrete (e.g., removing a specific connection from the graph). 

In this context, we need to design a semantic interaction model that interacts with the discrete nature of the higher-level structure layer in a meaningful way. To solve this problem, we propose a general framework that builds upon the Multi-Model Semantic Interaction (MSI) concept of Bradel et al. \cite{bradel2014multi} and addresses the challenges of developing SI pipelines with a mix of higher-level discrete structure space and low-level continuous models. Figure \ref{fig:interactions} shows an overview of the supported semantic interactions and the changes they induce on the extraction model.

In summary, the core contributions of this work are as follows:
\begin{itemize}
    \item The concept of the \textbf{Mixed Multi-Model Semantic Interaction (3MSI)} pipeline---an SI pipeline comprised of a high-level discrete structure and a lower-level continuous model that helps build the structure.
    \item An \textbf{SI Model for Narrative Extraction} that handles the intermediate abstraction layer defined by the narrative structure space and showcases the challenges of 3MSI.
    \item An \textbf{Evaluation} of the SI model to show whether it supports incremental formalism in the sensemaking process using a quantitative simulation-based approach and a qualitative approach with expert feedback.
\end{itemize}

\section{Background and Related Work}
%In this section, we discuss existing literature in the field of narrative extraction and visualization. Furthermore, we discuss works that use semantic interaction. 

\subsection{Narrative Extraction and Visualization}
Narratives are defined as systems of stories interrelated with coherent themes \cite{halverson2011master}. The same story can be told in countless ways, leading to a distinction between the underlying story itself and its representation. Narratological studies seek to understand the relation between stories and their representations \cite{puckett2016narrative, abbott2008cambridge}. 

Most computational narrative representation and extraction methods rely on event-based models \cite{shahaf2010connecting, liu2017growing, keith2020maps}. Events are the fundamental unit of narratives \cite{keith2022design} and they represent actions involving entities and characters or happenings without casually involved entities \cite{abbott2008cambridge}. However, event-based representations are not the only approach. For example, some works use a topic-level analysis, abstracting the narrative away from specific events \cite{nallapati2004event,kim2011topic,zhou2017survey}; some scholars propose even more fine-grained approaches, such as claims \cite{soni2014modeling} and named entities \cite{faloutsos2004fast}.

Furthermore, there are three general structures for narrative representations \cite{keith2022design}: timelines \cite{wang2015socially, shahaf2012connecting}, trees \cite{ansah2019graph, liu2017growing}, and directed acyclic graphs (DAGs) \cite{zhou2017emmbtt, shahaf2013information, keith2020maps}. Of these structures, DAGs provide the most flexibility by allowing the representation of divergent and convergent story substructures \cite{keith2020maps}.

Independent of the representational structure of the narrative model, extraction methods usually rely on optimizing a notion of narrative quality to generate the narrative representation \cite{keith2022design}. There are multiple optimization criteria, such as topical cohesion \cite{wang2015socially} (measuring whether two events share the same topic), coherence \cite{shahaf2010connecting} (measuring whether it makes sense to join two events together), and coverage \cite{shahaf2012trains} (measuring whether the narrative is properly covering the events). In this work, we use an extraction approach that maximizes coherence subject to coverage constraints \cite{keith2020maps}.

In the context of visual analytics and information visualization, narratives and storytelling are common techniques used to present data \cite{segel2010narrative,tong2018storytelling,riche2018data}. In particular, presenting visualization as stories can be used to aid in the sensemaking process \cite{hullman2013deeper, hullman2017finding}. Visual storytelling systems can help users detect relationships, structures, and other patterns, which can help in confirming hypotheses or gaining additional knowledge \cite{akaishi2007narrative,tong2018storytelling}. In this regard, constructing and interacting with visual narratives could be interpreted through the lens of a visual knowledge generation process \cite{sacha2014knowledge}, as users generate new knowledge as they work through their hypotheses and insights with the visual narrative model.

In this work, we develop an interactive visualization model of narratives based on the narrative map as defined by Keith and Mitra \cite{keith2020maps}---a DAG describing the connections between events (see Figure \ref{fig:covid} for examples). The original narrative extraction method for narrative maps is grounded in narrative theory and previous work has found that it provides analysts with a useful narrative representation for sensemaking tasks \cite{keith2020maps, keith2021vis}. However, subsequent work showed that this narrative representation could be improved by following a series of design guidelines \cite{keith2022design} based on an empirical study of how analysts create and use narrative maps to solve sensemaking tasks. Thus, we implement an improved version of the extraction method following these design guidelines to support our semantic interaction model. However, we note that the core contribution of our work is the semantic interaction model, rather than the incremental improvements to the narrative extraction process.

\subsection{Semantic Interaction}
Semantic interactions \cite{endert2012sisensemaking} exploit the natural interactions within a visualization---usually a projection of the data in a lower dimension space---to learn the intent of the analyst. In particular, instead of trying to manually modify parameters to model a specific concept, semantic interactions learn the parameters associated with that concept from user interactions and changes in the visualization. Semantic interactions have seen applications in text analytics \cite{endert2012semantic}, images \cite{hodas2016adding}, and other quantitative data with high dimensionality \cite{self2018observation}.

\begin{figure}[!htb]
    \centering
    \includegraphics[width=0.70\textwidth]{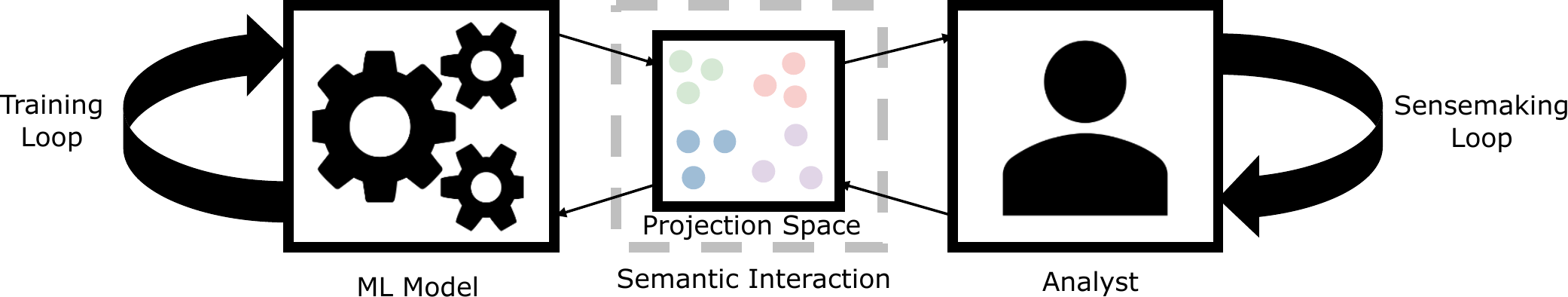}
    \caption{The traditional semantic interaction pipeline has the analyst interact with a dimensionality reduction model by making changes to the projection space. These changes are captured by the machine learning model and used to generate a new projection space. The process repeats as the analyst works through their sensemaking process.}
    \label{fig:si_pipeline}
\end{figure}

%In this context, SI systems can also be related to knowledge-assisted visualization systems \cite{chen2008data, federico2017role}, which rely on traditional knowledge bases to augment the capabilities of their models. These systems can capture user interactions to acquire tacit knowledge from the user and improve their own model \cite{lohfink2021knowledge}. However, unlike these systems, semantic interaction systems do not explicitly rely on external knowledge bases, but rather on machine learning models (e.g., dimensional reduction).

Figure \ref{fig:si_pipeline} shows semantic interaction as a bidirectional pipeline \cite{dowling2018bidirectional} based on an interactive DR model \cite{bradel2014multi}, the most common approach used when implementing SI systems. In this pipeline, analysts use the visualized projection to get insights, then they make changes to this projection by changing the position of data points. By making these changes, analysts are expressing their preferences. Thus, based on these changes, the interactive DR model is able to learn the intent of the analyst behind the changes in the projection. Using this information, the changes are converted back into a high-dimensional space. Afterward, the modified DR model updates the projection based on the new high-dimensional representations. Note that while this example is focused on DR, the underlying approach is not limited to just DR models and it can be generalized to other methods, such as force-directed graph layouts \cite{endert2012semantic, bradel2014multi}. 

Regardless of the implementation of the pipeline, it is necessary to capture the changes from the visualization and turn them into changes to the model. There are multiple machine learning models that attempt to solve the bidirectional transforms required to implement SI, such as Observation-Level Interaction \cite{endert2011observation}, Bayesian Visual Analytics \cite{house2015bayesian}, and Visual to Parametric Interaction \cite{leman2013visual}. Furthermore, the semantic interaction pipeline can be extended to leverage multiple models by chaining them and providing functionality and interactions at each level through the concept of Multi-Model Semantic Interaction \cite{bradel2014multi}. In general, any number of models can be used in this pipeline. Then, the interaction feedback from the user is interpreted as a change into one (or many) of the models using an appropriate inverse. These works show how DR models can capture the different interactions and modifications made by analysts. 

However, unlike the previous examples, narratives have an underlying temporal and causal structure \cite{abbott2008cambridge}, and in particular, graph-based representations of narratives have a discrete structure that has no natural continuous notion of distance that can be leveraged to define the necessary inverse transformation like in DR. Therefore, SI models for narrative or story visualizations must account for the underlying temporal structure and the discrete nature of the narrative representation, introducing an additional layer of complexity to the development of an SI model for narrative sensemaking. To address these issues, we develop the concept of 3MSI, which accounts for the usage of mixed models (i.e., a high-level discrete structure and a low-level continuous space) in the multi-level pipeline. Finally, to date, no previous research has sought to use SI in a narrative or story visualization setting. Thus, this research would provide a first step towards creating a human-AI interaction system that aids in the narrative sensemaking process by allowing the analysts to manipulate a narrative structure. %Moreover, by leveraging more abstract discrete structures beyond the projection space via the 3MSI framework, the proposed approach shows how to extend the semantic interaction pipeline with a mix of continuous spaces and high-level discrete structures. 

\section{Narrative Map Extraction}
\begin{figure}[!htb]
    \centering
    \includegraphics[width=0.75\textwidth]{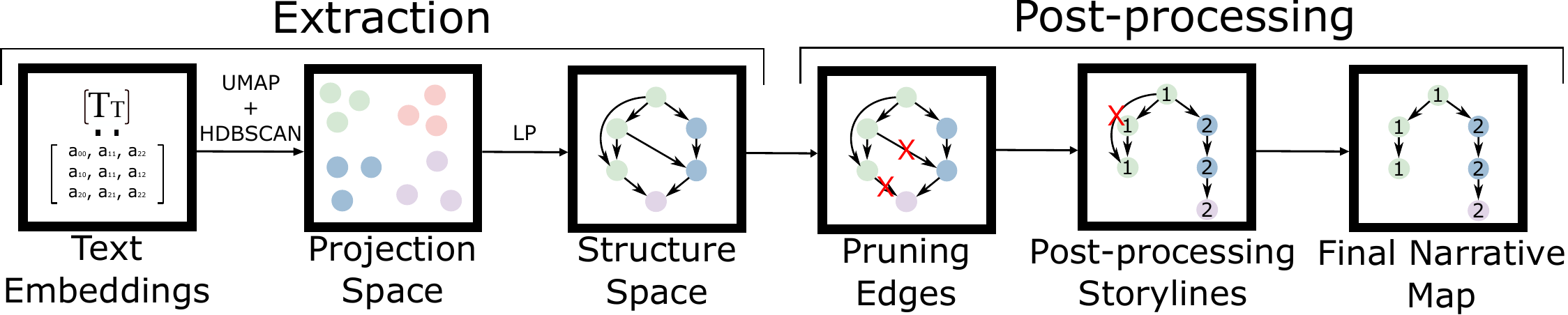}
    \caption{Narrative maps extraction pipeline. The pipeline consists of two main phases (extraction and post-processing).}
    \label{fig:map_extraction}
\end{figure}

In this section, we present our narrative extraction pipeline, which builds upon the extraction algorithm for narrative maps proposed by Keith and Mitra \cite{keith2020maps}. We further introduce optimizations to reduce computational costs and post-processing to align our results with the design guidelines for narrative maps \cite{keith2022design}. However, we note that the core contribution of our work is building an SI model for narrative maps, rather than the extraction process itself. We show the narrative extraction pipeline in Figure \ref{fig:map_extraction}. There are 2 phases in this pipeline: extraction and post-processing. 

The extraction phase builds upon the original extraction algorithm proposed by Keith and Mitra \cite{keith2020maps}, using UMAP to generate a projection space \cite{mcinnes2018umap}, HDBSCAN to compute topical clusters \cite{mcinnes2017hdbscan}, and linear programming to find the optimal narrative map. The narrative map corresponds to the structure space of our SI model. 

Next, the post-processing phase takes the basic map and simplifies it, following the design guidelines defined by Keith et al. \cite{keith2022design}. In particular, the post-processing phase seeks to decrease the overall complexity of the base map in order to reduce the cognitive load of the users. Specifically, post-processing involves removing low coherence edges, transitive connections inside each storyline, and redundant inter-story connections.

We note that there are two key steps in the extraction phase: the coherence computation step---which provides a quantitative measure of how much sense it makes to connect specific events---and the linear program (LP) formulation itself---which seeks to find the optimal narrative structure. We focus on these two steps to develop our SI model.

The coherence computation step relies on DR and clustering. We use the UMAP \cite{mcinnes2018umap} and HDBSCAN \cite{mcinnes2017hdbscan} algorithms, respectively. The coherence computation step is highly dependent on the underlying projection space. Thus, our SI model intervenes at the projection space level to induce changes in the coherence values. The LP formulation step handles structural constraints and integrates the coherence values from the previous step. Furthermore, the LP formulation can be modified to handle constraints induced by the user interactions directly. 

\begin{figure}[!htb]
    \centering
    \includegraphics[width=0.45\textwidth]{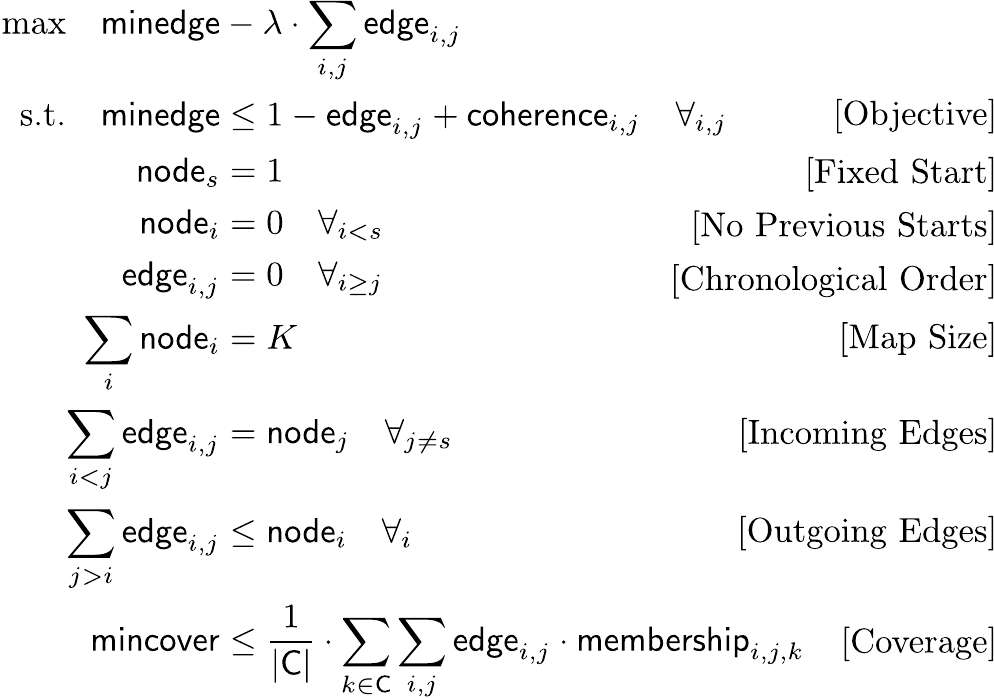}
    \caption{Linear program used to obtain the optimal narrative map (i.e., the basic structure space).}
    \label{fig:extraction_lp}
\end{figure}

We present the LP used for extraction in Figure \ref{fig:extraction_lp}, which seeks to maximize the coherence of the map. We note that compared to the original formulation \cite{keith2020maps}, our approach allows multiple endings and removes unnecessary constraints to reduce computational costs while maintaining the general structure. We provide more details in Appendix \ref{appendix:extraction}

Finally, we note that the interactive and iterative nature of SI can lead to overfitting issues \cite{choi2019concept} as the model attempts to satisfy the incremental requirements imposed by the user through each iteration of semantic interactions. Furthermore, each interaction performed by the user only affects a small subset of the data, which can lead to further overfitting issues, similar to how few-shot learning methods are prone to overfitting issues in general \cite{wang2020generalizing}. Thus, we include a regularization term that seeks to minimize the sum of all edge weights (i.e., $L_1$ regularization) and produce a sparse and less complex solution \cite{donoho2005sparse}. We discuss the effects of regularization in more detail in Appendix \ref{appendix:reg}.

\section{Semantic Interaction Model}
\label{sec:si}
In this section, we present our general 3MSI concept and how to address the challenges of using a mix of continuous spaces and discrete structures in the SI pipeline. Then, we present the SI model for narrative maps. %We start with an overview of our design goals that guide our model. %In particular, we show how we implement SI by integrating LP constraints into the extraction model and how we can use SI to change the results of the DR step of the narrative extraction pipeline. 

\subsection{Mixed Multi-Model Semantic Interaction}
The concept of 3MSI deals with a specific type of MSI where the top-level model of the SI pipeline---which is associated with the visualization shown to the user---is represented by a discrete structure space. Furthermore, the model must not have a continuous notion of distance that can be leveraged to model user interactions. For example, a graph represented with a force-directed layout can use a continuous notion of distance to model interactions, but a graph represented with a hierarchical layout would not be able to directly translate display distances into model changes (e.g., narrative maps). 

\begin{figure}[!htb]
    \centering
    \includegraphics[width=0.75\textwidth]{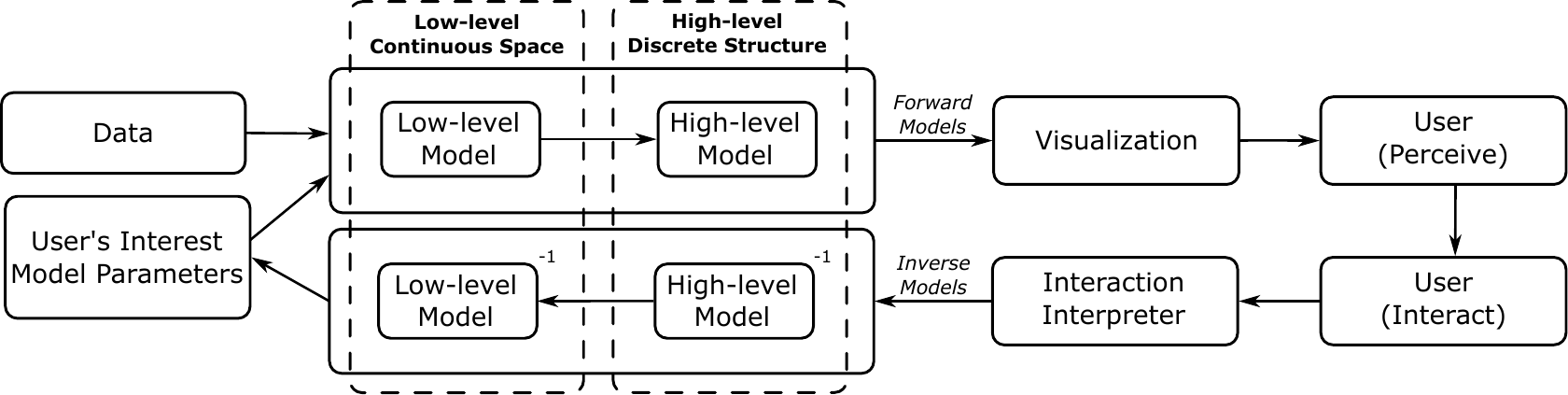}
    \caption{Generalized Mixed Multi-Model Semantic Interaction (3MSI) visualization pipeline.}
    \label{fig:3msi}
\end{figure}

The general 3MSI pipeline is shown in Figure \ref{fig:3msi}. In this pipeline, the lower-level model uses a continuous representation and the high-level model corresponds to a discrete structure. This model extends the traditional MSI approach \cite{bradel2014multi}---which only considers continuous spaces as internal models---with an additional layer of abstraction in the form of a discrete structure space, which provides users with further scaffolding to perform sensemaking tasks. 

Once the user perceives the visualization associated with the high-level model, they can interact with it. However, the introduction of a discrete structure in the 3MSI framework makes capturing user interactions and feedback for semantic interaction purposes more difficult compared to the traditional pipeline. The interaction feedback needs to be interpreted and fed to the inverse models in an appropriate manner. Defining how to interpret and feed the interactions back to the inverse models is the key step in using the 3MSI framework. Once the feedback has been captured by the models, the updated parameters are stored and used alongside the original data to update the visualization. This process repeats as the user keeps interacting with the updated models.

In the context of our work, the forward models are defined by the narrative extraction pipeline, with the low-level model corresponding to the UMAP projection and the higher-level model corresponding to the narrative structure itself. The inverse models depend on the specific type of interactions that we define and can rely on mathematically rigorous, heuristic, or probabilistic approaches \cite{dowling2018bidirectional}. In our implementation of 3MSI for narrative maps, we consider heuristic approaches, rather than defining formal mathematical inverses. In particular, we rely on adding specific constraints to the LP definition to modify the structure space and using semi-supervised learning to modify the projection space.

As mentioned before, interpreting the interactions and feeding them into the inverse models is the key step in the 3MSI pipeline. In particular, there are two key challenges that arise at this point. These two challenges are further amplified by the mixed nature of the pipeline: 
\begin{itemize}
\item The \textit{placement challenge}---how do we choose the levels at which we will handle each one of the relevant semantic interactions? 
\item The \textit{transformation challenge}---how do we transform changes from the higher-level discrete structure (e.g., the structure space) back to the low-level continuous space (e.g., the projection space)?
\end{itemize}

We note that these challenges are not necessarily unique to the mixed model context, but the discrete nature of the high-level model induces extra difficulties (e.g., defining an inverse function between the discrete structure and the low-level model). Furthermore, the solutions to these challenges depend on the specific architecture of the system. In this context, we propose general guidelines to aid in solving these challenges. In particular, we focus on providing ways to solve the placement challenge and circumventing the transformation challenge when possible, as this is the more complex challenge due to the difficulty of defining proper inverse transformations for these interactions, which can rely on mathematically rigorous, heuristic, or probabilistic approaches \cite{dowling2018bidirectional}. 

The proposed guidelines follow the general design principle of \textit{separation of concerns} and apply a \textit{divide-and-conquer} strategy when required for more complex interactions, adapting the general tenets of algorithmic and software design \cite{mens2012complexity} to the context of our SI framework. More specifically, the goal is to minimize the need of dealing with multiple levels at the same time---avoiding the transformation challenge. Assuming that the relevant semantic interactions have been identified, we present the general guidelines and specific examples pertaining to our narrative extraction case.

\paragraph{Handling Single-level Interactions: Separation of Concerns} 
As our core design guideline is to avoid the transformation challenge and reduce the implementation complexity of the 3MSI pipeline, interactions that have an intuitive and direct effect on a level should be implemented exclusively at that level and avoid changing the other level. This is the simplest case to implement for a semantic interaction, as it only requires interpreting the changes on a single level. For example, removing an event from a narrative map has a direct effect on the structure space (i.e., the model must create a representation that avoids that event). Thus, we implement this by only inducing changes to the structure space, without changing the underlying projection space, avoiding the transformation challenge.

Furthermore, interactions that can be solved at different levels in equally valid ways should be solved at the highest level possible (i.e., closest to the visualization and the user). As the model moves away from the visualization and deeper into the semantic interaction levels, each step makes the resulting changes potentially more opaque to the user. Thus, we recommend choosing the highest level possible, as this should provide a solution that is closer to the changes expected by the user, which would likely be biased by the visualization. In particular, this would be particularly relevant in cases where there are more internal lower-level models. In the narrative extraction context, removing an edge from a narrative map could be solved at the structure space level by simply imposing a specific constraint to avoid this edge in subsequent solutions or at the projection space level by forcing the events of that edge to be sufficiently distant to reduce the likelihood of being connected. However, following the guideline of using the highest level possible, we should choose the structure space level, as it is closer to the user and thus more likely to produce the intended effect.

\paragraph{Handling Multi-level Interactions: Divide and Conquer} Interactions that are more abstract and complex and do not have a direct effect on a single level should be separated into their constituent effects and each one of these should be implemented separately at a different level as appropriate. By separating the interaction into simpler components, we address both the placement challenge and the transformation challenge. Instead of relying on an explicit inverse function to get from the discrete structure space back to the continuous space or vice versa---which might not even be properly defined---we directly induce changes at each separate level following heuristics. This strategy can also be easily scaled to pipelines with many internal lower-level models.

For example, in contrast to the remove event example, grouping events is a more abstract and indirect interaction, as it does not specify how the events should be connected in the structure space. Instead, it simply implies that the events should be ``close'' and that the events should be somehow connected. Thus, we implement this interaction at both levels, inducing a change in the projection space (closeness) and adding constraints to the structure space (connectedness). We note that if it is not possible to adequately divide the semantic interaction by level, then we would have to deal with finding an appropriate inverse function instead of simpler heuristics.

\subsection{Semantic Interaction for Narrative Maps}
\subsubsection{Pipeline Overview}
Following the 3MSI concept, our approach has an additional abstraction layer compared to traditional SI systems: the structure space. This space builds upon the projection space, as it uses the projected representations to compute a discrete structure (the narrative map) that has no natural notion of continuous distance. The structure space captures higher-level relationships that might not be apparent in the projection space. 

In contrast to the traditional pipeline of Figure \ref{fig:si_pipeline}, the proposed pipeline has an additional level of abstraction: the structure space, which builds upon the projection space. The analyst interacts directly with the structure space (i.e., the narrative map), rather than with the continuous projection space. The changes in the structure space are used to update the underlying models, either by defining constraints that affect the narrative extraction process or by inducing changes into the projection space through semi-supervised learning. Thus, the semantic interactions with the narrative map can lead to either direct changes to the structure space or indirect changes to the projection space. 

\subsubsection{Design Goals for SI in Narrative Maps}
Following previous work \cite{keith2022design} and preliminary evaluations with users, there are two general ways in which narrative maps are used: \textbf{exploratory analysis}---where users are trying to find the different storylines in the data and understand the big picture---and \textbf{directed analysis}---where users seek to understand specific connections between events and stories. Thus, we seek to design an interactive model to support these analyses. In particular, there are three core subtasks that we support:

\textbf{Correcting Storylines:} Analysts need to be able to make corrections to existing storylines. This includes modifying intra-story connections (e.g., changing a specific edge, removing an event from the storyline) and inter-story connections (e.g., adding a connection with a different storyline to highlight common events or entities). This is mostly related to the directed analysis task, as users are refining the narrative schema.

\textbf{Creating Storylines:} Analysts need to be able to create new storylines and integrate them into the narrative map. For example, analysts might want to add new events or specify which events should be grouped into a coherent and consistent storyline. This is mostly an exploratory analysis task, as users are uncovering the different stories in the data.

\textbf{Shifting Focus:} Analysts need to be able to change the focus of the narrative map (e.g., changing the main storyline or the overall contents of the data). This interaction is relevant to both tasks, as users perform work through the sensemaking process they might require changing the focus of the narrative.

\subsubsection{Relevant Interactions}
We capture five relevant user interactions in our model. Examples of these interactions and their effects are shown in Figure \ref{fig:interactions}. We note that all the user interactions are done at the structure level, but they might imply a change in the underlying projection space too. Following the 3MSI framework, we need to identify the level at which these interactions must be handled and if necessary divide them into their constituent effects. We briefly describe the interactions and how they influence the underlying projection and structure space. %We show examples of these interactions and their usage in the case studies of Section \ref{sec:results}.

\textbf{Add/Remove Event:} Adding an event implies that the analyst would prefer for this specific event to appear in all subsequently extracted narrative maps. Likewise, removing an event implies it should not appear in any of the subsequently extracted narrative maps. In our implementation, these interactions only affect the model at the structure space level, without changing the underlying projection space.

\textbf{Add/Remove Connection:} Adding a connection between two event nodes implies that the analyst would prefer for this specific connection to appear in all subsequently extracted narrative maps. Likewise, removing a connection implies that it should not appear. These interactions only affect the structure space.

\textbf{Group Events (Clustering):} This interaction implies that the analyst would like for these events to appear in the same storyline or be close to each other in the narrative map (i.e., in the same route or at least connected in some way). As this interaction is more abstract and there is no clear definition of how a cluster should be handled in the structure space, we will handle this interaction at both levels (projection and structure space).

\subsection{Integrating SI into Narrative Extraction}
There are two steps in the extraction pipeline where we can implement semantic interaction: the coherence computation step (i.e., the projection space) and the linear programming step (i.e., the structure space). For the coherence computation step, we can modify the DR model used to generate the projection based on user interactions (e.g., forcing two events to be close together in the projection). For the LP step, we can modify the problem formulation by adding explicit constraints that force the solution to include the feedback from the user interactions (e.g., removing an event from the map should prevent it from appearing in the solution again). 

\begin{figure}[!htb]
    \centering
    \includegraphics[width=0.90\textwidth]{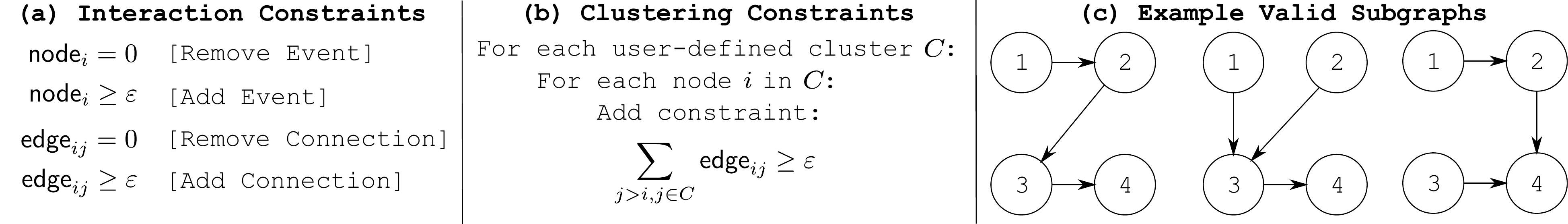}
    \caption{(a) Constraints induced by basic interactions (addition and removal). (b) Additional constraints to ensure that user-defined clusters are connected. (b) Examples of weakly connected subgraphs produced by the previous constraints in the narrative map.}
    \label{fig:weakly}
\end{figure}

\textbf{Single-level Semantic Interactions:}
We first deal with semantic interactions that can be handled at a single level by adding constraints directly to the LP formulation shown in Figure \ref{fig:extraction_lp}. This is a direct and straightforward method to model simple changes to the narrative map. We summarize all induced constraints in Figure \ref{fig:weakly}(a). In particular, node and edge removal can be directly incorporated by adding constraints of the form $\mathsf{node}_{i} = 0$ and $\mathsf{edge}_{ij} = 0$, respectively. These constraints ensure that nodes and edges removed by the user are not included in any of the subsequently extracted narrative maps. In contrast, node and edge addition are not as direct. It is not possible to include constraints of the form $\mathsf{node}_{i} > 0$ and $\mathsf{edge}_{ij} > 0$, as linear programming does not support constraints with strict inequalities \cite{zopounidis2002multi}. Instead, we use non-strict inequality constraints with a small positive number $\varepsilon$ to ensure that nodes and edges added by the user are included in all subsequently extracted narrative maps. We set $\varepsilon$ empirically, testing different threshold values we found that $0.01$ for edges and $0.05$ for nodes ensured that the elements were added in a mostly coherent way to the map.

We note that this approach only requires working on the structure space, as the constraints only induce changes to the extracted narrative map, rather than the lower-level projection space. Furthermore, as the user interacts with the visualization through their sensemaking process, the LP formulation will accumulate a series of induced constraints, forcing the optimization algorithm to find ways to fit the user feedback into the overall narrative structure it extracts.

\textbf{Multi-level Semantic Interactions:}
Now, we turn our attention to the more complex interaction of grouping events. Analysts can group events by manually selecting event nodes and assigning them a cluster number. For example, in Figure \ref{fig:interactions}, events $b$, $f$, and $h$ are assigned to Cluster 1. Unlike the previous cases, clustering events is a more complex interaction, as it not only implies that events should be connected or added directly in the structure space but also imposes the notion of these events being part of the same group (i.e., relatively similar).

Following the 3MSI guidelines, we divide the interaction into two effects: the events should be close in the projection space (closeness) and the events should be connected in the structure space (connectedness). Thus, we first use semi-supervised DR with the cluster information to keep the events close in the projection space. Next, we impose constraints on the LP formulation, seeking to keep the events connected in the structure space. This way, we induce changes in both levels of the SI pipeline without explicitly computing an inverse function between the structure space and the projection space. Instead, we rely on intuitive notions of closeness and connectedness to create appropriate heuristics. 

In more detail, we incorporate these manual clusters as training information for the DR step that uses UMAP. These clusters would be created based on user-defined labels (e.g., the user could assign all nodes with a specific keyword to be part of the same cluster) and fed to the extraction pipeline. UMAP supports semi-supervised learning \cite{mcinnes2018umap}. Thus, including the label information is direct and does not require further modifications. In particular, we create a label vector that contains the value $-1$ for all data points, which means that no labels have been assigned yet. Then, for each data point that belongs to a user-defined cluster, we assign the corresponding integer label to their entry in the label vector, following the format defined by the semi-supervised UMAP implementation \cite{mcinnes2018umap}. Next, we recompute the projection space using the label vector and the original embeddings. Note that when there are no user-defined clusters, we simply use the unsupervised version of the method. Once we obtain the new projection, we use it to recompute the similarity and clustering tables for the extraction algorithm.

Regarding the structure space changes, we add constraints to the LP to ensure that the clustered events appear on the map ($\mathsf{node}_{i} \geq \varepsilon$). We also impose further constraints---shown in Figure \ref{fig:weakly}(b)---by requiring the user-created cluster to be (weakly) connected, as shown in Figure \ref{fig:weakly}(c). As with the previous constraints, we set $\varepsilon$ empirically. We found that $0.01$ for nodes and $0.05$ for the edge sum worked well in most test runs. In conjunction with the new projection space, this ensures that the resulting map properly connects the grouped events. With all these changes, we solve the linear program again to find the new optimal narrative map.

%Finally, we note that this approach is similar to interactive clustering methods \cite{andrienko2009interactive,wenskovitch2017observation}. These methods allow users to impose constraints on whether specific data points should be in the same cluster or not \cite{wenskovitch2019pollux}. However, unlike these approaches, we do not apply these requirements at a cluster level. Instead, we rely on semi-supervised DR to encode this information in the projection space and on explicit constraints to alter the structure space.

\section{Evaluation Methodology}
\subsection{Data Set}
To show the effectiveness of our SI model we use news data covering the 2021 Cuban Protests that occurred in July 2021---the biggest protests in decades in Cuba \cite{robles2021protests}---which provide a sensemaking task of moderate difficulty for our experiments. We focus mostly on breaking news, as each news article represents a single main event \cite{norambuenaevaluating}. We scraped a data set of 500 online news articles from 20 news sources from different sides of the political spectrum. We categorized the sources based on external bias ratings taken from \texttt{AllSides.com}\footnote{allsides.com/media-bias/media-bias-ratings} and the Media Bias/Fact Check database \footnote{mediabiasfactcheck.com/search}.

Throughout our evaluation tasks, we rely on keywords and political leanings as labels. Previous research has shown that it is possible to computationally distinguish political leanings from the content of a news article \cite{karamshuk2016identifying, baly-etal-2020-detect}. Particularly, in times of political crises, there is a strong use of partisan content frames, which can be computationally detected \cite{karamshuk2016identifying}. Thus, we expect our SI model to be capable of learning such distinctions. 

%Furthermore, we note that these labels represent groups of events or connections based on user-defined criteria (e.g., all events containing a specific keyword or all connections between sources from different sides of the political spectrum). In practice, an analyst might not be interested in the specific labels used in our simulation, but it serves as a proof of concept on whether the semantic interactions are actually capable of learning relatively complex concepts.

\subsection{Simulation Tasks}
The human-centered approach \cite{boukhelifa2018evaluation} is the primary evaluation method for SI systems \cite{bian2020evaluating}. However, this method is highly dependent on human feedback and makes it challenging to compare different SI systems. This is due to the inherent difficulty of replicating user interactions in SI systems, as the semantic interactions build upon each other and modify the internal model through incremental formalism \cite{shipman1999formalism}. To deal with these issues, simulation-based evaluations \cite{bian2020evaluating,bian2021deepsi} work by creating a simulated analyst agent that attempts to replicate human interactions. Since there are no ground truths available for analyst intent and their cognitive process in general, these approaches use keywords or other easy-to-define criteria to label the data. Thus, this ground truth is used to guide the simulated interactions and compute error rates, allowing us to evaluate the ability of these models to perform incremental inference \cite{bian2020evaluating}. 

\subsubsection{Task Overview and Definitions}
We define five simple tasks with basic narrative goals and easy-to-define ground truths. These tasks use a series of labels that can be easily extracted from the data to simulate the creation of user-defined target labels (i.e., relevant events, consistent connections, etc.). These user-defined labels are based on keywords or other metadata that can be easily retrieved and do not require complex computations. In practice, users would define more complex and nuanced labels in their sensemaking process. Nonetheless, simple labels---based on keywords or pre-defined classifications---are sufficient to showcase the learning capabilities of our SI model. 

Furthermore, we note that the simulated analyst will only interact with a subset of the potentially relevant nodes or edges from the data set (i.e., those that appear on the map). Moreover, the model is not aware of these labels, it only knows about the embedding representation of the articles and the interaction feedback from the analyst. We expect the model to be able to \textit{learn} these user-defined labels based on the simulated analyst interactions. 

Moreover, we make the simulated analysts follow the same rules in each iteration (i.e., selecting events or edges based on a specific label). However, we note that each execution of the simulation is done using a different random seed, which leads to different narrative maps and, in turn, a different sequence of interactions in each task throughout the iterations, even if the simulated analysts always follow the same rules. We note that while it would be possible to create simulated analysts with more diverse interactions, such as interacting with random subsets of map elements or choosing random interactions, we opt to define simple rules for our simulation in order to reduce sources of variability. 

We note that the evaluation metrics all measure the \textit{error rate} of the model with respect to a specific element of interest. In particular, each task has a different metric based on its goal and the user-defined labels that we use (e.g., a specific set of relevant nodes, consistent connections, or consistently connected clusters). The tasks and their evaluation metrics are shown in Table \ref{tab:tasks}. We provide more details about the implementation of the simulation in Appendix \ref{appendix:simulation}.

%Finally, we note that these definitions could be modified and adjusted to fit other sensemaking tasks and types of interactions.

% \usepackage{booktabs}
\begin{table*}[!htb]
\centering
\resizebox{\textwidth}{!}{%
\begin{tabular}{@{}p{2.7cm}p{7.0cm}p{2.7cm}p{7.0cm}@{}}
\toprule
\textbf{Task}                   & \textbf{Description}                                                                                 & \textbf{Interactions} & \textbf{Evaluation Metric}                                                                         \\ \midrule
Remove Irrelevant Events (T1)        & Generate a narrative that does not include events corresponding to user-defined labels. & Remove Node                   & Fraction of \textbf{irrelevant nodes} with respect to the total number of nodes in the map.                                                    \\ \midrule
Remove Inconsistent Connections (T2) & Generate a narrative with consistent edges according to user-defined labels.              & Remove Edge                   & Fraction of \textbf{inconsistent edges} with respect to the total number of edges in the map.                                            \\ \midrule
Clean Up Storylines (T3)            & Generate a narrative with consistent storylines according to user-defined labels.         & Remove Node, Add Edge        & Fraction of \textbf{inconsistent nodes} with respect to the total number of nodes in the map.                                 \\ \midrule
Cluster Events (T4)                  & Generate a narrative with consistent clusters according to user-defined labels.           & Clustering                     & Fraction of \textbf{relevant nodes that are not connected} over the total number of relevant nodes in the map.    \\ \midrule
Add Relevant Events (T5)             & Generate a narrative that incorporates additional events in a consistent manner according to user-defined labels.           & Add Node, Clustering          &  Fraction of \textbf{relevant nodes that are not connected} over the total number of relevant nodes in the map.\\ \bottomrule
\end{tabular}%
}
\caption{Evaluation task description and evaluation metrics.}
\label{tab:tasks}
\end{table*}

\subsubsection{Specific Task Definitions}
\paragraph{T1 - Remove Irrelevant Events} We define three sub-tasks for T1 using different user-defined labels of increasing complexity. We start with a simple label based on keywords, where events are marked as irrelevant if they contain ``Florida'' or ``Miami'' in their headline (41 events). The goal is to create a narrative map that does not cover these specific events, as they are not relevant to the protests in Cuba itself or the US response in general. Next, we move to a more complex label based on the publication source of the event, where events are marked as irrelevant if they were published by Breitbart or Fox News (155 events). Both of these news sources are right-leaning and in general present a highly biased version of the events with right-wing framing. Thus, the goal is to create a narrative map that avoids these highly biased articles, thus providing a more neutral view of the narrative. Unlike the previous keyword-based approach, this requires the model to understand the differences between these highly biased sources and other sources. Finally, the last label is also based on the publication source of the event, where events are marked as irrelevant if they were published by any right-leaning news outlet (207 events). Generating a map that excludes this label is even harder than the previous example, as it considers highly biased sources the same as sources with only mild bias, making the distinction fuzzier.

\paragraph{T2 - Remove Inconsistent Connections} T2 is based on our definition of inconsistent edges (connecting right-leaning to left-leaning articles and vice versa). Intuitively, minimizing inconsistent edges in a narrative map leads to a map that does not have abrupt changes in the framing and coverage perspective. Thus, the goal of removing these inconsistent edges is to create a map that avoids drastic changes in the framing and presentation of the facts. We seek to create a map that does not directly connect left-leaning and right-leaning articles. Thus, any change in the political leaning of a storyline would be mediated by an unbiased article, allowing for a slower shift in framing.

\paragraph{T3 - Clean Up Storylines} T3 is based on our definition of consistent nodes (i.e., having the same political leaning as their storyline). The goal of cleaning up storylines is to create internally consistent storylines that share their political leaning and avoid storylines that shift their political leaning. This is similar to the goal of removing inconsistent edges, but using a mix of node and edge operations at a storyline level, rather than edge operations on the whole map. Ideally, this would lead to a map with parallel storylines showing how each side of the political spectrum presents the events, allowing analysts to contrast these different perspectives.  

\paragraph{T4 - Cluster Events}
The goal of this task is to create a narrative map that properly covers the events from the user-defined clusters, by creating a consistent presentation of these events. That is, the nodes of these clusters should be connected according to our definition. The SI model should be capable of inferring other events (modeled by the changes in the projection space) that should be in the cluster and connecting them properly, even if the user did not explicitly label them as part of the cluster. We define two sub-tasks for T4.

For the first sub-task, we define two clusters based on headline keywords. The first cluster (41 articles) is made from articles focusing on Florida and Miami. The second cluster (96 articles) is made from US-focused articles, excluding Florida and Miami. The goal is to create a narrative map that includes coverage of the general US response and the additional protests in Florida. %Note that due to the definition of our evaluation metric, it does not suffice to have the relevant articles sprinkled throughout the narrative map, instead, they have to be properly connected. 
For the second sub-task, we define two clusters based on publication source and political leaning. The first cluster (83 articles) is made exclusively from articles published by Breitbart, a highly biased news source. The second cluster (92 articles) is made from articles published by unbiased sources (AP news and Reuters). The goal in this example is to create a narrative map that includes highly-biased coverage from Breitbart and unbiased coverage. The resulting map would help an analyst compare and contrast these perspectives.

\paragraph{T5 - Add Relevant Events} We define two sub-tasks for T5. For the first sub-task, we define relevant events as those that contain ``Florida'' or ``Miami'' in their headline (41 articles). The goal is to generate a narrative that includes events about the protests in Florida. To do this, the analyst would mark relevant events in the map as part of a single cluster and add more relevant events. These two interactions tell the narrative extraction model that the relevant events are closely related and that the narrative map should also integrate more relevant elements into the map, respectively. For the second sub-task,  we define relevant events as those that contain ``Biden'' in their headline (92 articles). The goal is to generate a narrative map that includes events about the response of the US president to the Cuban protests.

\subsection{Qualitative Evaluation}
To perform a qualitative evaluation of our model, we first implemented an interactive prototype that allows analysts to extract narrative maps and use semantic interactions on the extracted narrative maps. Using this prototype, we studied two different usage scenarios for SI and narrative maps (COVID-19 and Cuban protests). Then, to evaluate whether our proposed SI model provides value to actual users, we performed an evaluation and review with three experts in visual analytics working in the intelligence analysis domain. Furthermore, the prototype was shown to two analysts working in the same domain. In particular, we discussed the general capabilities of the model and the prototype, the resulting maps of each case study, and the effects of the semantic interactions as we navigated the different examples. Due to space constraints, we only present the COVID-19 case study in the main article and leave the other case in Appendix \ref{appendix:case}.

\textbf{Case Study.}
We used a data set about news on COVID-19 taken from the design guidelines study \cite{keith2022design} which contains 40 documents. The goal of the analyst in this case study is to understand the global spread of COVID-19 and its effects during January 2020. We note that, unlike the simulated analysts, we do not attempt to correct the map until its error rate is zero in the case study. Instead, we show a more natural approach that combines different types of semantic interactions to evaluate their quality as an analysis tool in a more realistic context. Furthermore, we only perform a few iterations, showing that the model provides valuable insights even with only a limited amount of semantic interactions. 

\begin{figure*}[!htb]
    \centering
    \includegraphics[width=0.95\textwidth]{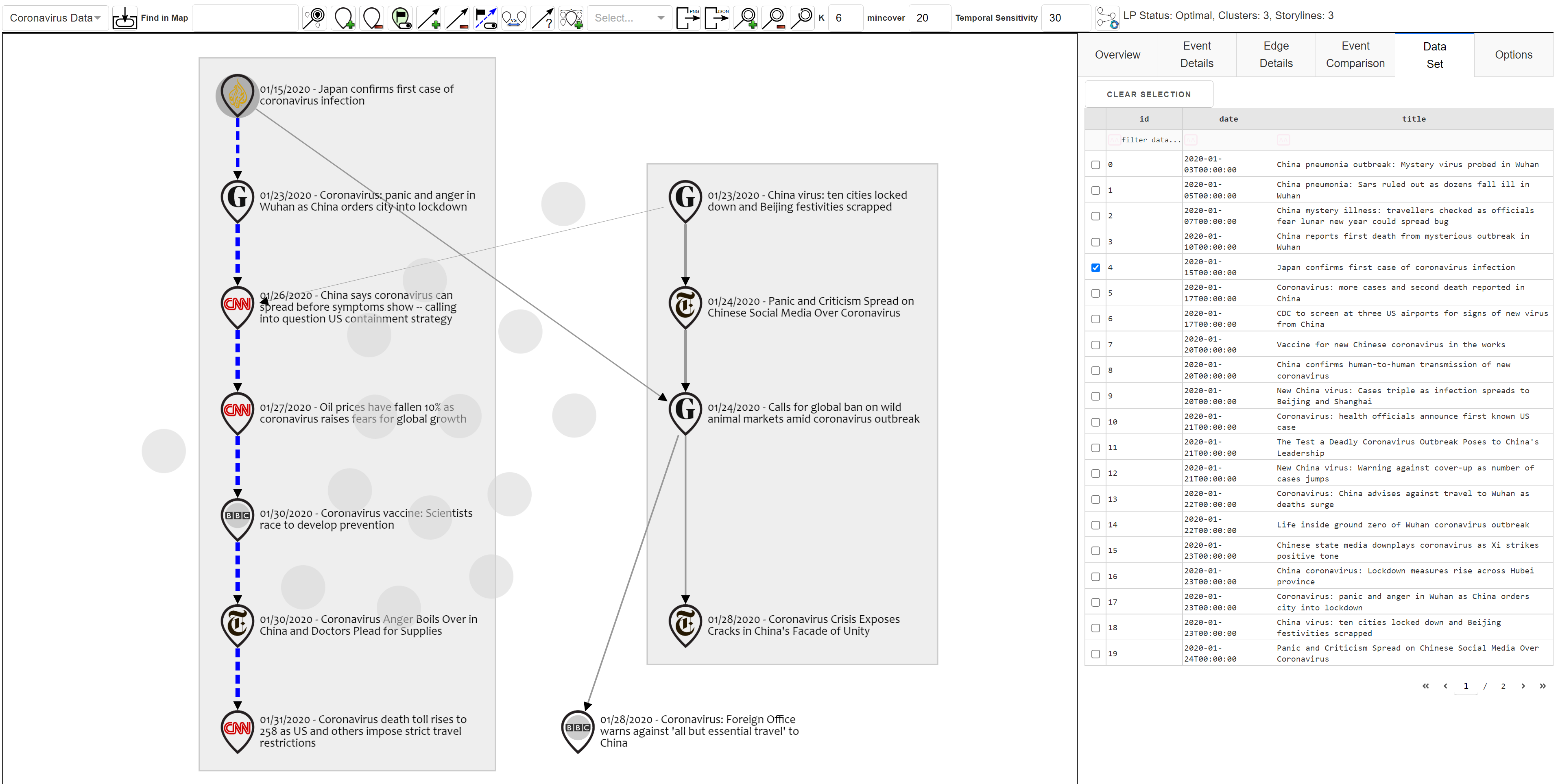}
    \caption{Prototype interface showing an example extracted map in the main canvas, the data set table in the right tab, and the main menu and extraction parameters on the top.}
    \label{fig:interface}
\end{figure*}

\textbf{Interactive Prototype.}
We implemented an interactive prototype using the Dash Cytoscape library \footnote{https://dash.plot.ly/cytoscape}. The graph layout of the narrative map is generated using GraphViz \cite{ellson2004graphviz} and the DOT language \cite{gansner1993technique}. We show the interface in Figure \ref{fig:interface}. This interface supports all the interactions and tasks defined in our simulation methodology and case studies, as well as basic navigation and providing details-on-demand about the elements of the map. 

The prototype shows the narrative map on the main canvas on the left. Events are shown on the map with the landmark icon \resizebox{0.25cm}{!}{\includegraphics{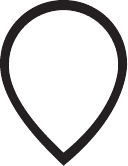}} and connected with edges \resizebox{0.30cm}{!}{\includegraphics{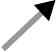}}. Nodes include the logo of the news outlet that published the corresponding article. The main storyline is shown with the dashed blue line \resizebox{0.30cm}{!}{\includegraphics{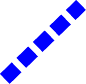}}. Edge width is based on the coherence value of the connection.

We note the addition of background gray nodes \resizebox{0.30cm}{!}{\includegraphics{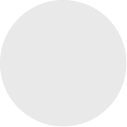}} that show similar events to those in the map but were not chosen by the optimization algorithm (i.e., they do not improve the coherence of the map). These extra nodes are useful to guide the ``Add Event'' interaction, although that interaction can also be done directly from the data table. The gray nodes are organized using a force-directed layout based on similarity. We note that the original narrative maps implementation did not show related events \cite{keith2020maps}, focusing only on the extracted structure. However, we chose to display them to provide analysts with potentially relevant events and help with the node addition interaction.

The top menu allows selecting the data set and loading it into the system \resizebox{0.40cm}{!}{\includegraphics{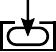}}. The top menu also contains the core interactions that allow users to manipulate the narrative map, such as ``Add Event'' \resizebox{0.40cm}{!}{\includegraphics{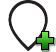}}, ``Remove Event'' \resizebox{0.40cm}{!}{\includegraphics{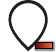}}, ``Add Connection'' \resizebox{0.40cm}{!}{\includegraphics{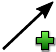}}, ``Remove Connection'' \resizebox{0.40cm}{!}{\includegraphics{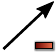}}, and ``Cluster Events'' \resizebox{0.40cm}{!}{\includegraphics{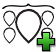}}. The parameters of the extraction algorithm (expected length $K$, minimum average coverage, and temporal sensitivity $\sigma_t$) can also be modified by the user. To create a map, the user must click the ``Generate Map'' button \resizebox{0.40cm}{!}{\includegraphics{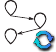}}. 

The right panel has six tabs containing additional information and options (overview, event details, edge details, event comparison, data set details, and additional options). For the purposes of the case study, we only used the event details tab (i.e., the contents of the article) and the data set details tab (i.e., a list of all the articles), as the rest are used for development purposes. Finally, we note that the prototype contains other supporting interactions that can aid in navigation and analysis (e.g., searching), but these interactions are not used in the current SI model.

\section{Results}
\label{sec:results}
%We now present our results. In particular, we start with our simulation results for all the defined simulation tasks, showcasing the capabilities of the SI model. Next, we present our qualitative analysis of two case studies, followed by our discussion with expert users and the analysis of their feedback.

\subsection{Simulation Results}
\begin{figure*}[!htb]
    \centering
    \includegraphics[width=0.85\textwidth]{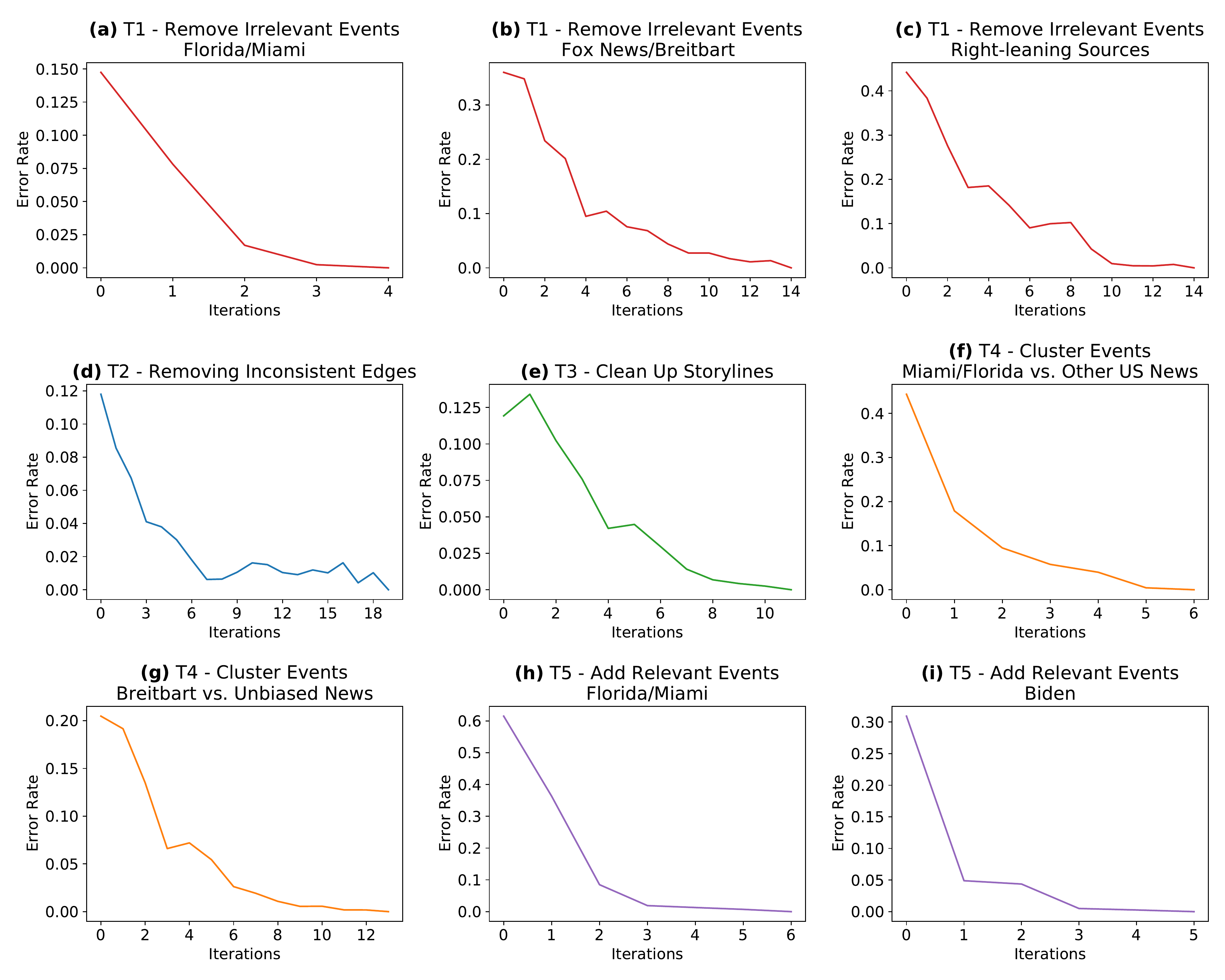}
    \caption{Error rate vs. the number of iterations. \textbf{(a, b, c)} T1: Remove Irrelevant Events. \textbf{(d)} T2: Remove Inconsistent Edges. \textbf{(e)} T3: Clean Up Storylines. \textbf{(f, g)} T4: Cluster Events. \textbf{(h, i)} T5: Add Events.}
    \label{fig:results}
\end{figure*}

We now present our simulation simulation-based evaluation results for each task, including the average error rate in each iteration and how many iterations are needed in each task to achieve the target error rate. Figure \ref{fig:results} shows all the simulation results. We note that for all tasks, the error rate tends to zero after a sufficient number of iterations. Moreover, in most cases, it only takes a few iterations to reach a low error rate (e.g. around 5\% error rate). However, some tasks are harder than others, taking more iterations to converge. %In some cases, even inside the same task, we see some variability in the convergence rate.

The hardest task---in terms of iterations needed---corresponds to the edge-based task T2. In particular, after the initial drop in error rate, the number of inconsistent edges stabilizes below 2\%, but it takes up to 18 iterations to completely remove them. We note that this is an inherently harder task compared to event-based tasks due to the number of possible combinations of edges that can arise in a graph, which rise quadratically with the number of nodes. Nevertheless, if our goal is to remove most of the inconsistent edges and achieve a mostly consistent narrative map, we could stop between 7 to 9 iterations, which results in around 2\% inconsistent edges. 

As an example of intra-task variability, consider T1, where the second and third cases take longer than the simpler case of removing Florida/Miami events. The slower convergence is expected, as these two specific sub-tasks are more complex, as the user-defined label does not rely on simple keywords, but rather the abstract concept of political bias. Likewise, T4 shows similar behavior, where the keyword-based clustering creates an easier task compared to the more abstract clustering based on bias vs. unbiased news.

\subsection{Case Study}
\begin{figure*}[!htb]
    \centering
    \includegraphics[width=0.98\textwidth]{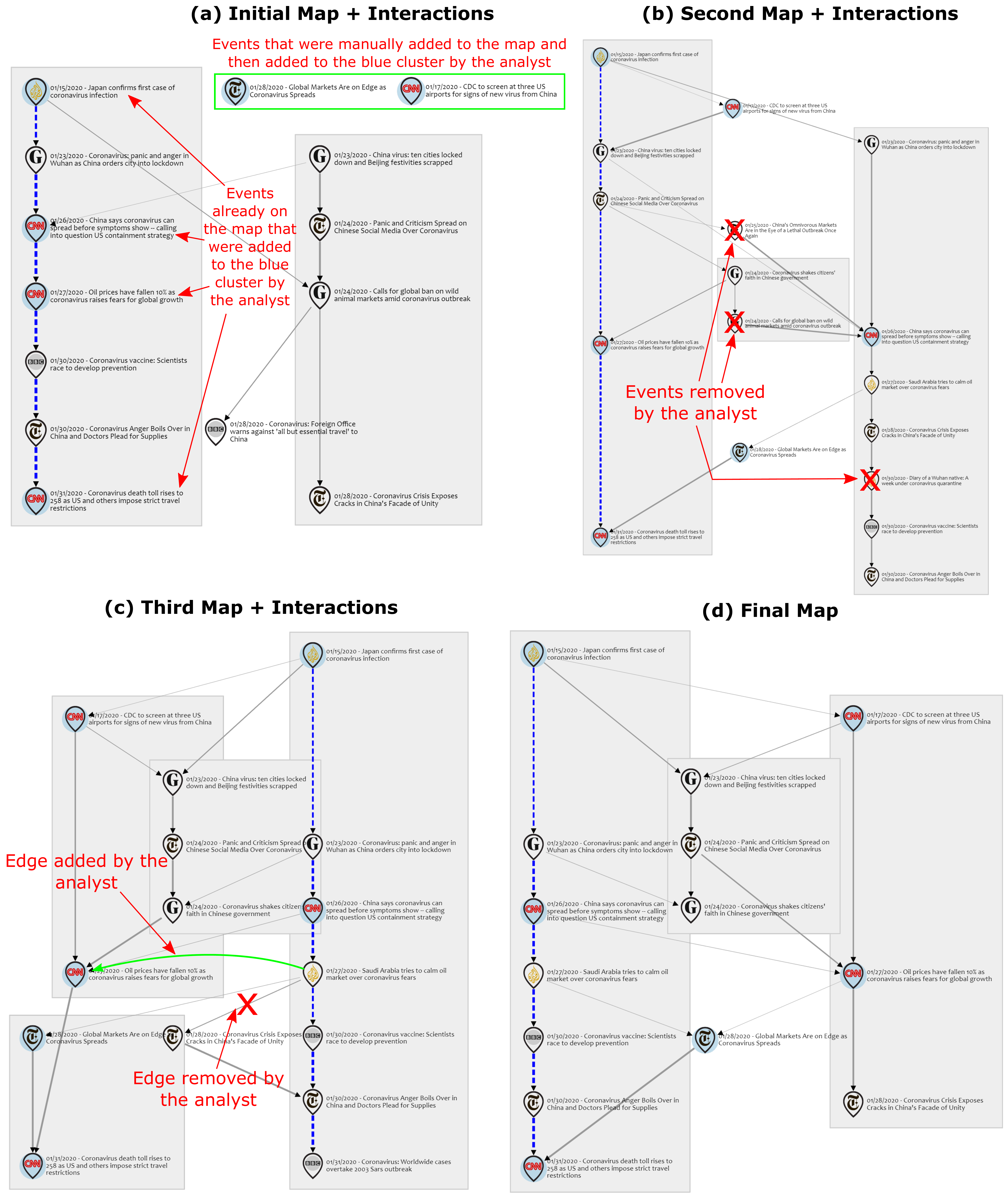}
    \caption{(a) Initial map and analyst interactions (add events and cluster events). (b) Second map and analyst interactions (remove events). (c) Third map and analyst interactions (add edge and remove edge). (d) Final map after all iterations. Note that a higher-resolution version is available as part of the supplementary materials.}
    \label{fig:covid}
\end{figure*}

We now present the qualitative analysis of the COVID-19 case study as an extended example. The goal of the analyst in this case study is to understand the global spread of COVID-19 and its effects during its first month in January 2020. However, during this time period, most of the news articles reported events from China, leading to the data set being mostly focused on Chinese news. Thus, the analyst needs to set an appropriate starting point that is more likely to lead to a map detailing the global spread of the virus and its effects. In this context, the analyst sets the 5th event of the data set---which reports Japan's first COVID-19 case---as the starting point, as it corresponds to the first news article that is not about China in the data set. In terms of parameters, the analyst uses $K = 6$ for the map size and default values for the rest. Figure \ref{fig:covid} shows all the maps created throughout the analysis process and the interactions done by the analyst.

Figure \ref{fig:covid}(a) shows the initial map and the interactions of the analyst. The map has two major storylines---shown in the map as linear subgraphs in gray boxes. The main storyline represents the most important path in the narrative map and is computed using maximum likelihood \cite{keith2020maps} and is shown with blue edges. All other connections use gray edges. Finally, there is also a singleton storyline---a storyline with only one element---shown without a gray box. 

On a more detailed inspection, the main storyline has some relevant events about the effects of the virus and its spread (e.g., comments about the US containment strategy, oil prices, vaccines, and travel restrictions). However, there are also some events about the effects in China that are not necessarily useful for the analyst's goal. The other storylines are also all focused on China. Thus, to shift the focus of the map towards the global spread and its effects, the analyst performs the sequence of interactions shown in Figure \ref{fig:covid}(a). 

In particular, the analyst adds two potentially relevant events to the map: global markets being on edge and the CDC screening at airports. The first event relates to the global economic effects and the second to the international prevention strategies. Afterward, to ensure that the added events are integrated consistently into the map, the analyst clusters them together with other relevant events from the main storyline (see the highlighted events in Figure \ref{fig:covid}(a)).

Then, after completing the interactions, the analyst generates the new map shown in Figure \ref{fig:covid}(b). This map still has two storylines, and the events highlighted in blue that we assigned to the same cluster are mostly connected (some indirectly). However, there are still some issues, such as irrelevant events and the fact that the map still focuses on China. To fix the first issue, the analyst removes the irrelevant events on animal or omnivorous markets---which are potentially important as a cause of the start of the pandemic, but not relevant to its ongoing spread and effects---and the event about daily life under quarantine---which is also irrelevant for the analyst's goal. 

These interactions lead to the map shown in Figure \ref{fig:covid}(c). This map has four major storylines and all highlighted events are connected (technically, their induced subgraph is weakly connected). The main storyline now covers more relevant events, and although it still contains events about China, they fit into the overall context (e.g., lockdowns and lack of supplies). Two of the storylines consist of only highlighted events, thus they focus exclusively on the global spread and effects on a worldwide scale. Furthermore, these storylines are connected through relevant inter-story connections with the main storyline. Furthermore, the map now has distilled an extra storyline about Chinese socio-political effects into its own storyline. Thus, providing a better structure to represent the narrative.

While the current map provides a good overview, it could be improved further. In particular, the analyst could add a direct edge between the two events about the oil market. Furthermore, the analyst could disconnect the event about Saudi Arabia from the Chinese facade of unity event, as they are unrelated. These operations lead to the map shown in Figure \ref{fig:covid}(d). This time the interactions did not introduce major changes. In fact, two storylines are preserved from the previous map (main story and Chinese socio-political effects). The third story is about relevant side effects, such as the US screening for the virus leading to oil prices dropping due to fear. However, the last event of this storyline does not seem to fit and could be removed to further improve the representation. The only singleton storyline of this map is placed in an interesting position, its connections imply that the fear in oil markets has extended to global markets. %This event then converges back to the main story into an event about the global travel restrictions, another reason why global markets were on edge.

Overall, the final structure showcases some of the key storylines about the early impact of COVID-19, including the global spread and the ensuing economic issues. Furthermore, the map still contains relevant information about the socio-political effects in China, which were not prevalent in other countries at this point (e.g., lockdowns and unrest). All these storylines provide valuable insights into the pandemic and even foreshadow future effects of the pandemic.

\subsection{Expert Feedback}
%We reviewed the prototype alongside extended examples of interactions with three experts in visual analytics working in the intelligence analysis domain. Furthermore, the prototype was shown to two analysts working in the same domain. In particular, we discussed the general capabilities of the model, the resulting maps of each case study, and the effects of the semantic interactions as we navigated the different examples, and general feedback about the prototype.

First, the experts provided feedback on the general narrative maps framework, highlighting how the prototype could provide analytic value by generating a structured representation of the documents organized over time into different storylines. Second, regarding the SI model, the experts found that it was able to properly capture user feedback in most cases and examples, leading to improved maps or clearer storylines. Furthermore, we discussed the potential value of the model for exploratory tasks and directed tasks. For exploratory tasks, the value of the model was clearer, as they provide analysts with a series of potentially relevant storylines, which can be then explored in-depth via SI. In contrast, for directed tasks, the value would depend on the type of objective, as not all directed tasks would need a structured approach (e.g., they could be solved with simpler search-based models). However, if the task required finding complex connections, then the narrative model provided more value.

In general, the experts found the general narrative maps framework useful and the SI model capable of properly capturing user feedback. However, they highlighted two potential issues: \textit{transparency} and \textit{scalability}. In particular, the current SI model acts as a black box, and understanding its effects is not straightforward. Thus, it could be complemented with an explainable AI approach to provide clear explanations of the changes induced by the model. Regarding scalability and performance issues, we note that this is one of the main drawbacks of the current implementation, as the optimization approach and the layout algorithm have issues scaling to larger data sets. Nevertheless, the experts noted that scalability and transparency are generally considered open problems in complex sensemaking systems. Thus, regardless of these issues, the experts found the 3MSI narrative model to be a welcome advancement to help during the sensemaking process.  Finally, the experts noted that they had not seen any previous work that integrated SI capabilities with narrative extraction, which provided additional value and it would be ``extremely useful'' for certain sensemaking tasks.

\section{Discussion}
%We now discuss our results and the limitations of our methods.

\subsection{Semantic Interaction Model}
Our results show that the proposed 3MSI system is capable of integrating analyst feedback into the narrative model. In all tasks, the average error rate tends to zero as the analyst refines the model, although we note that the convergence rate depends on the specific task and its inherent difficulty. Thus, our experimental results support the hypothesis that our 3MSI model for narrative maps is effective at supporting incremental formalism \cite{shipman1999formalism} based on iterative interactions. 

In general, our 3MSI concept provides a flexible approach that can enable researchers to use higher-level discrete structures as outputs of a semantic interaction pipeline in conjunction with a lower-level continuous model (or even multiple lower-level models). Furthermore, our proposed guidelines for handling single-level and multi-level interactions can aid researchers and practitioners in the construction of complex SI models. However, while the 3MSI approach is flexible, it could become increasingly complex depending on the number of levels we include in the pipeline. %The deeper the pipeline, the more difficult it would be to address the placement and transformation challenges, as each level would require its own interaction model. 

Furthermore, in the context of interactive AI \cite{wenskovitch2020interactive}, if we seek to add an explainable AI component to complement our SI model, each additional layer in the system's architecture increases the complexity of generating explanations, as each transformation between layers requires some way to ``invert'' it and extract meaningful information. This increasing complexity makes dealing with the transparency issues highlighted in our expert review a more complicated endeavor. 

%Finally, we note that the implementation of the specific pipeline depends on the specific structure space and internal models used in the system. Researchers working on such systems would need to address the placement and transformation challenges according to their specific needs. The general guidelines proposed in our work could aid them in addressing these challenges. However, in general, developing a model-agnostic approach that does not require redesigning interactions for each specific case remains an open challenge that future work could address.

\subsection{Semantic Interactions and Narrative Sensemaking}
Our simulation-based analysis provides a replicable and scalable evaluation for our narrative SI models without requiring human analysts. However, we note that due to the relative simplicity of our test tasks compared to real sensemaking tasks, these results might overestimate the incremental formalism capabilities of the SI model, yet it provides a good baseline. In practice, we would expect real sensemaking tasks to require more iterations to refine the model. Nevertheless, the fact that, on average, we were able to learn from different types of analyst interactions is promising. 

Regarding our qualitative evaluation, our case studies and subsequent expert evaluation show that SI is capable of providing value to analysts and aiding in the sensemaking process. In particular, both the final structure provided by the narrative map after SI and the intermediate maps generated in each step are useful for gathering insights from the data. In this context, despite the main goal of semantic interaction being to generate a “good” final map by trying to capture user intent through successive iterations, the iterative process itself also has value to the analyst. Even if each individual iteration is not a perfect map, the exploratory nature of this process can provide valuable insights, which can be compounded as the analyst explores different configurations of the structure space. Even in cases where SI fails to generate a ``good'' map, there is value in the exploratory analysis that it allows.

Overall, our findings suggest that combining semantic interactions with narrative visualizations can assist in the sensemaking process and allow analysts to change the story according to their own preferences. Intelligence analysis \cite{endert2014human}, computational journalism \cite{bradel2015big}, and other fields that require analyzing and understanding narratives could benefit from our proposed SI model for narrative maps.

\subsection{Model Transparency and Performance}
Two of the key issues of the current SI model as identified by our analysis with experts are \textit{transparency} and \textit{scalability}.

Regarding transparency, the system should provide appropriate explanations of the changes it makes to the narrative to the users. Thus, it would be useful to complement SI with an explainable AI model that provides such feedback. In the context of creating a general interactive AI model \cite{wenskovitch2020interactive} for narrative maps, adding an explainable AI component to our pipeline would be the natural next step. Furthermore, we also note that the transparency issues are likely accentuated due to the usage of non-incremental layout algorithms for the output graphs, which can lead to additional cognitive load due to drastically different positioning of events.

%In particular, after each iteration of semantic interactions a whole new layout is computed, leading to potentially highly different positioning of the same nodes (see Figure \ref{fig:covid}). This potentially drastic change in layout brings a heavy cognitive load to the user observing the narrative maps and performing sensemaking with them. Hence, using an incremental layout algorithm \cite{incremental1996dag} and highlighting the specific changes made by the semantic interactions could aid in reducing cognitive load. Further possible improvements to help users understand the underlying model include providing in-depth explanations for connections and storylines.

Regarding performance and scalability, we note that it would be possible to reduce the computational cost by applying techniques to group documents together and treat them as a single event, rather than keeping the assumption of a one-to-one relationship between documents and events. Such an approach could be helpful to reduce the computational cost of the map extraction process and the cost of the subsequent layout generation. However, it would require extensive pre-processing and appropriate machine learning models to detect which documents refer to the same event. 

In this context, referring back to topic detection and tracking literature \cite{nallapati2004event, khurdiya2011multi} to identify events or other structured narrative extraction approaches \cite{tannier2013building, liu2017growing, ansah2019graph} that represent events as clusters of documents could be useful. In particular, adding an event extraction model in the middle of the proposed 3MSI pipeline for narrative maps could help reduce the overall computational cost and make the model more scalable, as the last step to generate the structure space is the most computationally expensive part of the pipeline.

\subsection{Interaction Ambiguity}
While our SI model was able to capture knowledge from the simulated analyst, there is ambiguity in how to interpret the user interactions properly. The challenge of defining user intent from interactions has been called the ``With Respect to What'' problem \cite{wenskovitch2020respect,hu2013semantics} in SI literature. %For instance, take the example in Figure \ref{fig:cuba}. After generating the new map, our model created side storylines related to the protests in Florida, as well as other relevant events, but the main storyline still focused on Cuba. However, the analyst could have intended to entirely shift the focus of the map from the Cuban protests towards Florida or the US, rather than treating them as side storylines. 
Possible solutions to this issue include providing additional interactions \cite{wenskovitch2020respect} so that the user can specify with more precision what they actually want to obtain (e.g., by using additional interactions) or providing the ability to control the impact \cite{hu2013semantics} of the interactions (e.g., by designing a weighting scheme that integrates elements from original representation and the post-interaction representation). In our particular example, analysts could specify that the current stories should be kept together by using the clustering interaction. By doing so, the model would be forced to keep the original storylines in the representation while also integrating the new story. Thus, ensuring that the model properly captures the incremental formalization process, rather than throwing away the previous narrative map. However, further exploration is required to define a robust model of user interaction for narrative maps that captures user intent correctly when there is ambiguity.

Furthermore, based on the feedback from the experts, an important step that can aid in understanding how the interactions are working is explicitly showing the changes introduced by the semantic interactions. Adding such explanations would help users determine whether the interactions are capturing their intentions properly when there is potential ambiguity. Furthermore, in case the semantic interactions have misinterpreted the intent of the analyst, they could apply corrective interactions to the highlighted changes and guide the extraction algorithm to generate a new map that deviates less from the original intent.

\subsection{Limitations}
Our work is not without limitations. While the general 3MSI framework provides a general approach to semantic interaction, our specific SI approach for narrative maps is not model agnostic. The current model is built around the linear programming extraction model and the use of a DR method for the embedding representation of documents. Thus, any changes to the extraction pipeline would require figuring out how to integrate constraints into methods. 

Regarding the evaluation of the SI model, we note that the simulation-based experiments only used simple tasks that do not represent the full complexity range of narrative sensemaking tasks that analysts could perform with a narrative map. To ameliorate this issue, we presented in-depth case studies that attempt to replicate real sensemaking tasks and gathered feedback from experts in visual analytics working in the field of intelligence analysis. 

Finally, we note that there are other potential interactions that could be used to capture the intent of analysts. For example, manually changing the main story of the narrative map or highlighting a specific event to mark it as more important. However, the current model does not handle such interactions in its current form. Despite these limitations, our quantitative and qualitative results show that our 3MSI model has the potential to help analysts modify computational narrative models through simple interactions with narrative visualizations, without having to understand the underlying extraction or representation models used to generate the narrative visualization.

\section{Conclusions}
In this paper, we proposed the concept of 3MSI---Mixed Multi-Model Semantic Interaction---an SI pipeline defined by a higher-level discrete structure and lower-level continuous models. Furthermore, we showcased the capabilities of this approach by developing an SI model for narrative maps, which had a lower-level model in the form of a document embedding projection space and a higher-level narrative structure space. 

We evaluated our SI model through a quantitative simulation-based approach. The evaluation showed that the SI model is capable of integrating the simulated analyst's interactions into the narrative model. Thus, supporting the hypothesis that the SI models are effective at supporting incremental formalism for narrative maps. 

Furthermore, our review of case studies and validation with experts led to valuable feedback regarding the SI model in particular and the interactive narrative maps framework in general. The expert feedback showed that the SI model for narrative maps could provide value to analysts in their sensemaking process. In terms of broader impact, integrating SI models with narrative maps allows us to better support the sensemaking loop for narrative sensemaking tasks, such as journalistic analysis of news narratives \cite{bradel2015big} and intelligence analysis \cite{endert2014human}. However, we note that there are two key issues in the current SI model for narrative maps: \textit{scalability} and \textit{transparency}. Thus, future work should seek to improve the extraction approach to allow for the analysis of larger data sets and explainable AI techniques should be used to improve the transparency of the SI model and the changes it makes to the narrative maps. %Moreover, to further evaluate the capabilities of the model, future work should also include a larger-scale user study of the SI model and the interactive narrative maps prototype with more complex narrative sensemaking tasks than those used in our experiments.

Finally, in terms of broader impact, the 3MSI concept provides a flexible framework that allows researchers to employ higher-level discrete structures as outputs of an SI pipeline in combination with different continuous internal models and representations. Furthermore, our proposed guidelines for dealing with single-level and multi-level interactions can help academics and practitioners build complex SI models.

\section*{Acknowledgments}
This work was supported in part by NSF I/UCRC CNS-1822080 via the NSF Center for Space, High-performance, and Resilient Computing (SHREC), the NSF grants CNS-1915755 and DMS-1830501, and by ANID/Doctorado Becas Chile/2019 - 72200105.

\bibliographystyle{ACM-Reference-Format}
\bibliography{sample-base}

\appendix
\newpage
\section{Narrative Maps Extraction}
\label{appendix:extraction}
In this appendix, we provide more details about the narrative maps extraction pipeline, including the coherence computation step, the linear programming formulation, and the post-processing steps.

\subsection{Coherence Computation}
The coherence computation step relies on the projection space and the associated topical clusters \cite{keith2020maps}. Thus, it depends on the dimensionality reduction and clustering algorithms used to build this space. In particular, coherence is computed as the geometric mean between content similarity and topical similarity. The content similarity term is computed based on the normalized cosine similarity of the projected document embeddings, which are based on the \textsf{all-MiniLM-L6-v2} model of the \textsf{sentence-transformers} library \cite{reimers-2019-sentence-bert}. The topical similarity term is computed using the Jensen-Shannon similarity \cite{hadjila2019flexible} to compare the distribution of the topical cluster probabilities of each document.

However, this definition of coherence leads to issues when dealing with data sets that span longer periods of time, such as generating long-ranging connections between events that are seemingly related but too temporally distant to be relevant. To deal with this, previous approaches have used exponential decay factors based on the temporal distance between events \cite{zhu2012finding, liu2017growing, camacho2019analyzing, liu2020story, yu2021multi}. Thus, we extend the previous notion of coherence with a temporal component that penalizes events that are temporally distant using an exponential factor. Let $\Delta t_{i,j}$ be the temporal distance between events $i$ and $j$. The new value for coherence is defined as
\begin{equation*}
\mathsf{coherence}_{i,j} = \exp{\left(
-\frac{\Delta t_{i,j}}{\sigma_t}
\right)} \cdot \sqrt{\mathsf{content\mhyphen similarity}_{i,j} \cdot \mathsf{topical\mhyphen similarity}_{i,j}}.
\end{equation*}
where $\sigma_t$ represents the temporal sensitivity, a user-defined parameter for the extraction algorithm that regulates the rate of decay. For the purposes of our work, we measure all temporal distances and $\sigma_t$ in days and we use a default value of 30. However, for other data sets and contexts, it could be necessary to change the units of temporal measurement.

\subsection{Linear Program Formulation}
Now we provide more details about the components of the linear program formulation (see Figure \ref{fig:extraction_lp}).

\textbf{Variables and Parameters.}
The map is built by finding the optimal weights for its nodes and edges. These weights are represented by the $\mathsf{node}_{i}$ and $\mathsf{edge}_{i, j}$, respectively, where $i$ and $j$ represent the indices of the corresponding events in the data set. These variables take values between $0$ and $1$. There are three user-given parameters: $K$, the expected main story length that regulates the size of the map; $\mathsf{mincover}$, the minimum average coverage for the topical clusters; and $\lambda$, the strength of the regularization term in the objective function. There are also pre-defined parameters that depend on the results of the projection space. Namely, the coherence values of each connection $\mathsf{coherence}_{i,j}$, and the membership of an edge to a given topical cluster $\mathsf{membership}_{i, j, k}$ \cite{keith2020maps}. In this context, $C$ represents the set of topical clusters (indexed by $k$). Finally, the $\mathsf{minedge}$ variable represents the minimum coherence value of the narrative map over the edges and is used as the optimization goal. 

\textbf{Objective Function and Constraints.}
Following the model of Keith and Mitra \cite{keith2020maps}, the linear program seeks to maximize the value of the minimum edge, under the assumption that the best narrative structure should have no weak connections (i.e., the minimum coherence of the map is as high as possible). This is done using the $\mathsf{minedge}$ constraints that depend on the coherence values. 

In our evaluations, we fixed the regularization strength parameter ($\lambda$) to the inverse of the total number of potential edges in a directed acyclic graph of size $n$. That is, we set $\lambda = \frac{2}{n(n-1)}$. Our empirical evaluations showed that this value worked well for our purposes (see analysis of results in Appendix \ref{appendix:reg}).

\textbf{Starting Event Constraints.}
We note that the aforementioned design guidelines study showed that analysts tend to use maps with a single source event and potentially multiple endings \cite{keith2022design}. Thus, we allow users of our extraction algorithm to select a single starting event $s$ when extracting the map. This is handled by the $\mathsf{node}_s = 1$ which sets the corresponding starting node as active. Furthermore, to avoid interfering with the user-defined starting event, we prevent events that occur before the starting event from appearing on the map, which is handled by the constraint that assigns their node variables to $0$. We note that it would be possible to adapt this approach to a fixed ending event, or other combinations of starting and ending events, including none of them.

\textbf{Chronological Order.}
To ensure that the events in the map follow a chronological order, we impose a series of constraints based on the index of the events. This constraint assumes that the data set has been sorted by time before running the extraction algorithm.

\textbf{Map Size Constraint.}
This constraint determines the expected length of the main storyline on the map and thus it regulates the map size. This constraint is determined by the parameter $K$, which is defined by the user. We note that the original extraction model \cite{keith2020maps} had another constraint to ensure that the sum of edge weights added up to $K-1$. However, we removed this constraint and replaced it with the regularization term that seeks to minimize this sum.

\textbf{Edge Constraints.}
These constraints seek to generate the connection structures of the map, relating the incoming and outgoing edges of each node with the activation value of that node. Unlike the original formulation by Keith and Mitra \cite{keith2020maps}, we use inequality constraints for the outgoing edges, as this approach allows for maps with multiple endings.

\textbf{Coverage.}
To ensure appropriate topical diversity and coverage over the narrative map, the original formulation of the extraction algorithm also included a coverage constraint based on the average coverage of each topical cluster (obtained from the HDBSCAN clusters in the projection space). A topical cluster is considered to be covered if sufficient edges belonging to the cluster (defined by the $\mathsf{membership}_{i,j,k}$ parameter) are on the map. The minimum coverage required ($\mathsf{mincover}$) is a user-defined parameter. We set 20\% minimum coverage as the default value of this parameter.

\subsection{Post-processing}
We note that despite the use of regularization, the optimal map in terms of coherence could still be highly complex and have redundant edges or connections that make it difficult for a human to interpret adequately. Thus, as described in the extraction pipeline, we perform a series of post-processing on the optimal map for usability purposes and to keep in line with the design guidelines defined by Keith et al. \cite{keith2022design}. 

\textbf{Pruning Edges.}
In particular, after extracting the initial narrative map, we pruned edges to reduce the cognitive load for users. In particular, we only keep a certain number of edges per node, thus, restricting the branching factor of the DAG. Furthermore, after the initial pruning, we remove any leftover edges that have too low of a coherence value in the final model. 

More specifically, we only leave the top $\lceil\sqrt{K}\rceil$ outgoing edges of each node and remove the rest. Next, we removed all edges that had less than $0.1 / K$ coherence after re-normalizing edge weights. In both formulas, $K$ is the map size parameter of the linear program described in the previous subsection. These heuristics were found through empirical testing. We note that the cut-off values depend on $K$ because a fixed cut-off value became problematic for larger maps, as it leads to too many edges being filtered. These heuristics helped prevent most cases of excessive edge density without eliminating potentially relevant connections.

\textbf{Storyline Identification.}
Next, we also identified the different storylines of the narrative map. This is a necessary step to remove the transitive connections from storylines (i.e., redundant connections) and to minimize inter-story connections (i.e., only keep the first and last of such connections for each pair of stories). Furthermore, identifying these storylines is used for evaluation purposes (e.g., analyzing a specific storyline of a map). However, the original narrative map extraction method does not directly generate a partition of the map into storylines, it only highlights the main storyline \cite{keith2020maps}. Thus, we need to develop an algorithm to partition the map into storylines. 

To do this, we use the recursive algorithm described in Algorithm \ref{alg:recursive_algorithm}. This algorithm takes a narrative map as an input (i.e., a weighted directed acyclic graph), with potentially many sources (starting events) and sinks (ending events). This algorithm is based on the idea that we can recursively extract the storylines by finding the most coherent paths of the graph in each step. The first storyline is easy to identify, as that would simply be the main storyline, which we identify by finding the maximum likelihood chain of the graph based on coherence. To find the rest of the storylines, we first remove the events from the main storyline and all the corresponding connections. Then, we find the new ``main'' storyline using whatever events and connections remain in the graph. We recursively perform this process and store all the extracted storylines until only storylines of size 1 remain (i.e., a graph of disconnected single events). As we remove storylines, we might end up separating the graph into multiple connected components. Intuitively, these components correspond to different parts of the narrative (e.g., topics) and should contain different storylines.

\begin{algorithm}[!htb]
\caption{Recursive algorithm to extract all the storylines based on their overall coherence.}\label{alg:recursive_algorithm}
\SetAlgoLined
\DontPrintSemicolon
\KwIn{Narrative map $G = (V, E)$ - Weighted directed acyclic graph with weights between 0 and 1.}
\KwOut{List of sequences of events (storylines).}
\hrulefill
\\ 
\SetKwProg{Fn}{Function}{:}{}
\Fn{\GraphStories{G}}{
\If{$|E|$ == 0}{\Return \List{\Singletons{V}} \Comment*[r]{Base Case - graph with no edges.}}
$sp$ = \ShortestPath{G} \Comment*[r]{Shortest Path (Negative Log Likelihood).}
$H = G - sp$ \Comment*[r]{Delete Nodes in $sp$ (and edges) from graph $G$.}
$H$ = \Normalize{H} \Comment*[r]{Normalize outgoing edge weights (sum = 1).}
\Return \List{sp} + \GraphStories{H} \Comment*[r]{+ means list concatenation.}}
\end{algorithm}

\textbf{Transitive Reduction.} We note that the original implementation of the narrative map extraction algorithm had an additional constraint to reduce redundant transitive connections \cite{keith2020maps}. However, this approach introduced a cubic number of constraints in the worst case, greatly increasing the computational cost. Thus, removing this constraint from the optimization problem and performing post-processing to remove transitive connections reduces the overall computational cost of the extraction process, allowing a more responsive system.

In particular, for each identified storyline in the graph, we use transitive reduction \cite{aho1972transitive} on the corresponding storyline sub-graph making sure to keep the edges with higher coherence values. Afterward, we re-normalize the edges and obtain the final storyline. Note that this process does not affect inter-story connections.

\textbf{Inter-story Connections.}
For inter-story connections, we only keep the first and last of such connections for each pair of storylines on the map. To do this, we find the boundary edges between the sub-graphs defined by each storyline. Then, we keep the first identified connection and the last identified connection, based on the indices of the corresponding nodes. After removing all redundant inter-story connections, we re-normalize the edges of the graph.

\section{Simulated Analyst Implementation}
\label{appendix:simulation}
In this appendix, we describe our simulation experiment and the implementation of the simulated analyst. 

\subsection{Simulation Overview}
Throughout our experiments, we test the ability of the models to learn incrementally over the course of several interactions with the simulated analyst. Due to the computational costs of executing the narrative map extraction process and the subsequent interactions and refinements, we only take 10 samples of each task. Each sample will generate slightly different narrative maps, as the underlying embedding spaces are projected and clustered with a different seed each time. Our results show that regardless of the starting conditions, after sufficient interactions the models are capable of learning based on the simulated interactions. Furthermore, we fixed the starting event for all maps to reduce variability and make comparisons easier. In particular, we chose a news article from an unbiased source (Reuters) describing the start of the protests: ``Cuba sees biggest protests for decades as pandemic adds to woes''\footnote{https://www.reuters.com/world/americas/street-protests-break-out-cuba-2021-07-11/}. 

In each iteration, we first compute the evaluation metric for the current narrative map using the definitions in Table \ref{tab:tasks}. Then, the simulated analyst performs a series of interactions with the model and generates a new map. We rinse and repeat until the error rate achieves the desired target level (0\%). Finally, we discarded all samples that started with a 0\% error rate, as it would not make sense to use semantic interactions to improve the narrative model in those cases. We kept executing our simulation until we obtained 10 valid samples.

Finally, we note that the SI system has no awareness of the user-defined labels throughout the evaluation process. Only the simulated analyst knows these definitions and uses them to determine its actions in each task. Moreover, the analyst only interacts with a subset of the potentially relevant elements from the data set, but we expect the model to be able to \textit{generalize} based on these interactions. For example, for T1 with the Miami/Florida label (41 events in the data set), the average number of total user interactions throughout the simulation was approximately 7 node removals. Thus, the different metrics seek to capture the model's generalization ability.

\subsection{Simulation Definitions}
We provide more details on the definitions used for the task definitions and associated evaluation metrics in Table \ref{tab:tasks}. In particular, the tasks make use of the following definitions:

\begin{itemize}
    \item \textbf{Relevant and Irrelevant Nodes} We define event nodes as \textbf{relevant} if they are associated with a user-defined label, which can be defined by the presence of a keyword, the political leaning of the associated news outlet, or even the news outlet itself. We define \textbf{irrelevant} nodes in a similar manner.
    \item \textbf{Inconsistent Edges} We define edges as \textbf{inconsistent} if they connect articles from opposite political leanings. For example, articles from right-leaning outlets with articles from left-leaning outlets, or vice versa. Any other combination is considered consistent.
    \item \textbf{Inconsistent Nodes} We define event nodes as \textbf{inconsistent} with respect to their storyline if they conflict with the most frequent political leaning of their storyline, excluding unbiased articles. Ties are broken by selecting the first political leaning in the storyline.
    \item \textbf{Connected Nodes} We define event nodes as \textbf{connected} with respect to their cluster if they fulfill any of the following conditions based on general narrative structures \cite{keith2020maps}: they are directly connected to another relevant node of the same cluster, they are in the same storyline as another relevant node of the same cluster, or they share direct predecessors or successors with another relevant node of the same cluster. 
\end{itemize}

\subsection{Simulated Analyst Actions}

We now describe the actions taken by the simulated analyst in each iteration for each task. We note that each iteration consists of multiple actions (e.g., multiple node removals or additions).

\textbf{T1 - Remove Irrelevant Events.} 
The simulated analyst removes all nodes marked as irrelevant according to the defined target labels. These interactions should eventually lead to a narrative map that avoids events with this keyword. When there are no more events to remove from the map, the error rate is zero and the simulation ends. 

\textbf{T2 - Remove Inconsistent Connections.}
The simulated analyst removes all nodes marked as inconsistent according to our criteria. These interactions should eventually lead to a narrative map that avoids inconsistent connections. When there are no more connections to remove from the map, the error rate is zero and the simulation ends. 

\textbf{T3 - Clean Up Storylines.}
For each storyline in the narrative map, the simulated analyst removes all inconsistent edges. To do this, the simulated analyst first computes the political leaning of a storyline by selecting the most frequent political leaning, excluding center articles. The simulated analyst breaks ties by selecting the political leaning of the first biased article in the storyline. 

\begin{figure}[!htb]
    \centering
    \includegraphics[width=0.30\textwidth]{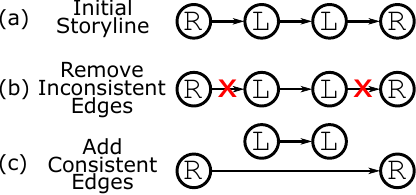}
    \caption{Example of the simulation steps for task T3 (Clean Up Storylines). (a) The initial storyline. (b) Removing the inconsistent edges from the storyline. (c) Reconnecting the storyline with consistent edges.}
    \label{fig:consistent_example}
\end{figure}

Once the political leaning of a story has been chosen, all nodes inside that have the opposite leaning are marked as inconsistent. To make the storyline consistent, the simulated analyst disconnects the inconsistent nodes from the rest of the storyline by removing the inconsistent edges in the storyline. Then, the simulated analyst reconnects the storyline by adding edges between the events that are missing connections. See Figure \ref{fig:consistent_example} for an example of this interaction. 

These interactions should eventually lead to a narrative map with only consistent storylines. Note that we do not delete the inconsistent nodes, as they could be re-used as part of another storyline by the extraction algorithm. When there are no more inconsistent nodes in all storylines of the map, the error rate is 0\% and the simulation ends. 

\textbf{T4 - Cluster Events.}
For each pre-defined cluster label, the simulated analyst marks the relevant nodes in the narrative map as part of their respective clusters. These interactions should eventually lead to a narrative map that integrates the relevant clusters as part of the map by properly connecting them in the same storyline or through common predecessors or successors. When there are no more relevant isolated nodes on the map, the error rate is zero and the simulation ends. The simulation also ends if there are no more relevant nodes to add to the cluster (note that this termination condition did not occur in practice during our experiments).

We note that our clustering task only considers disjoint clusters. The current implementation of our SI model is not able to handle the multi-label case, where a single event belongs to more than one cluster.

\textbf{T5 - Add Relevant Events.} 
The simulated analyst marks the relevant nodes present in the narrative map as part of a single cluster. Then, they add new nodes to the map by taking the first 10\% (rounded up) of the relevant events that did not appear on the map from the data set. 

In this task, we cannot measure success based on the simple frequency of relevant events associated with the user-defined label, because our goal is not to generate a map filled with only the relevant events. Instead, we care about the added events being properly integrated into the narrative map, which is why we need to use both clustering and node addition in this task. Thus, we use the same metric from the clustering task (T4), but with only a single ``cluster''. Thus, like in T4, when there are no more relevant isolated nodes on the map, the error rate is zero and the simulation ends. The simulation also ends if there are no more events to add (although this ending condition did not occur in practice). Preliminary evaluations showed that using node addition without clustering does not lead to convergence, because the model does not understand that the added nodes are necessarily related.

\section{Case Study: Cuban Protests}
\label{appendix:case}
\begin{figure*}[!htb]
    \centering
    \includegraphics[width=0.75\textwidth]{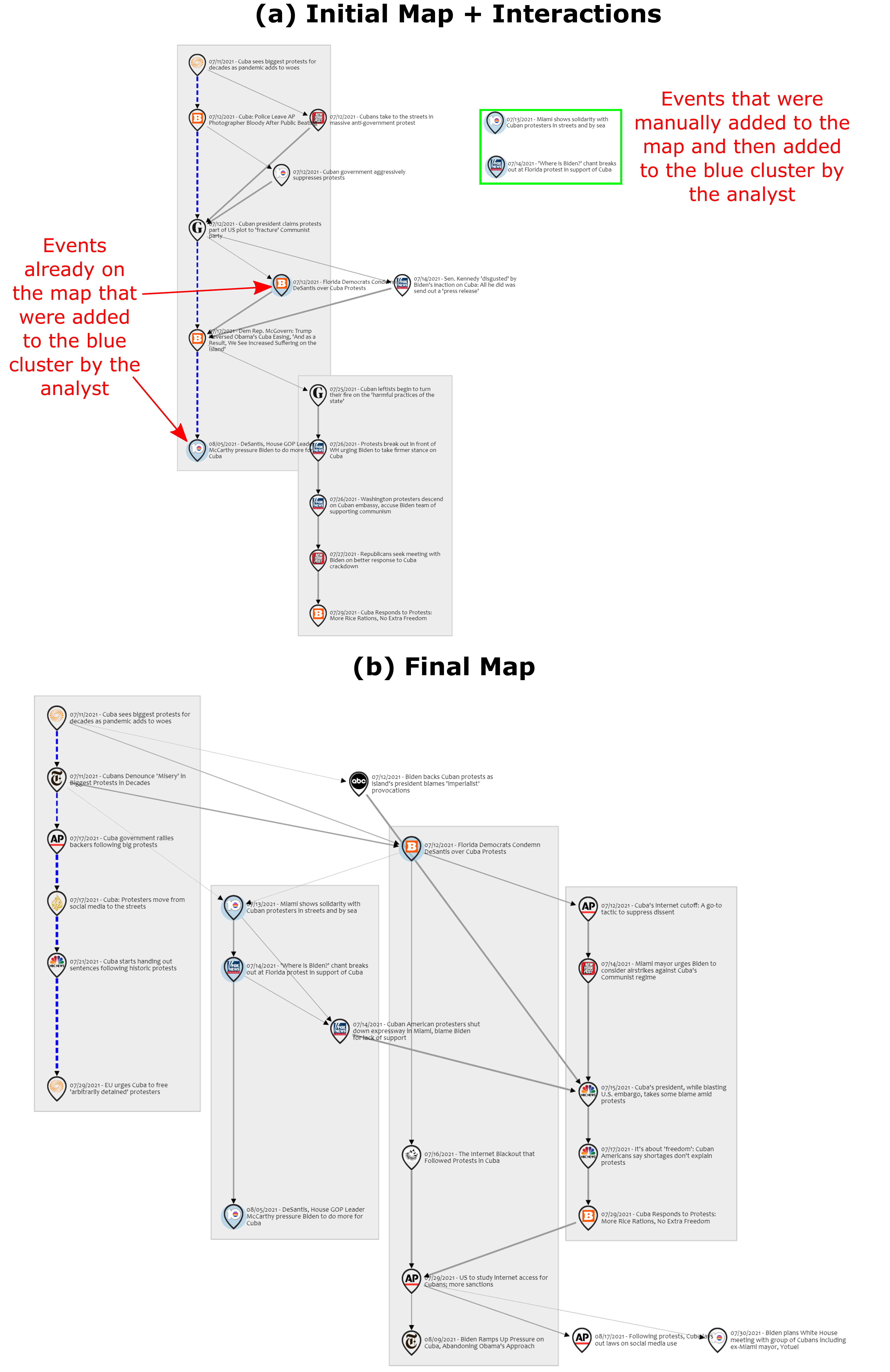}
    \caption{(a) Initial map and interactions performed by the analyst (add events and cluster events). (b) Final map after a single iteration. Note that a higher-resolution version is available as part of the supplementary materials.}
    \label{fig:cuba}
\end{figure*}

The goal of the analyst in this task is to gain an understanding of the causes and effects of the Cuban protests using a smaller version of the Cuban data set with only 160 documents. For the effects, the analyst considers both what happened in Cuba itself and any relevant developments in the US, due to its geopolitical importance in this context. Relevant effects range from the general response by the administration to more specific socio-political effects. The analyst generates a map using the same event we used for our simulation experiments, namely: ``Cuba sees biggest protests for decades as pandemic adds to woes''. In terms of parameters, the analyst uses $K = 6$ for the map size and default values for the rest. Figure \ref{fig:cuba} shows the resulting maps and interactions of the analyst.

Using these parameters, the system generates the map shown in Figure \ref{fig:cuba}(a). This map has two major storylines, as well as four minor singleton storylines. The main storyline starts focuses on the Cuban protests at first, but it slowly starts to drift toward US-related issues. The side storyline that comes afterward is focused mostly on socio-political effects in the US, including protests and political struggles. However, there are two events that are relevant to the Cuban protests themselves. Furthermore, the first two singleton storylines simply reinforce some aspects of the protests (e.g., aggressive suppression of the protests). The second two singleton storylines, which happen after the Cuban president claims that the protests are a US plot, provide some insights into how the US is responding to these events, particularly, there is political infighting in Florida and calls to action from the senate. Despite some useful insights, the current structure is not that practical as it has a disjointed coverage of different aspects of relevant events. Thus, the analyst will perform semantic interactions to generate a better structure to represent the narrative. 

In particular, the analyst wants to find out more about the effects on Florida, as several events mention this state and related entities or keywords (e.g., DeSantis and Miami). The analyst adds two relevant protest events, as seen in Figure \ref{fig:cuba}(a). Next, the analyst groups these events with the DeSantis events that were already present the clustering interaction. The expected result of this interaction is a map that focuses more on the effects on Florida and the US. 

The actual results are shown in Figure \ref{fig:cuba}(b), where we see four major storylines and four singleton storylines. The main storyline is now exclusively focused on the Cuban protests, including counter-protests and government repression. which is a nice side effect of the interactions. Second, we see a storyline that has only highlighted events that the analyst clustered, corresponding to a storyline focused on Miami. Next, the third storyline starts with a DeSantis event but then turns its attention to a series of events about trying to provide internet access to Cubans. This connection makes sense when delving deeper into the article's contents, which mentions that DeSantis is pushing for the administration to provide internet access to the island. One of the singleton storylines right after this also provides further insights, as Cuba outlaws social media usage after the protests. The last storyline also mentions the internet cutoff on the first event, but then shifts focus toward calls for military actions against the Cuban regime. Afterward, the storyline gives us some insights into the potential causes of the protests from different points of view. From the Cuban perspective, the US embargo was the key cause, although the government also admits some blame for the economic shortages and lack of supplies. From the Cuban Americans' perspective, it is more about freedom than supply shortages. Finally, the last event highlights how the final response of the government was to attempt to fix the supply issues by giving more rations without any fundamental change to the regime. While further improvements would be possible, the current layout and contents provide a decent overview of the main developments of the Cuban protests, and the protests in Florida, as well as some insight into the potential causes of the protests (e.g., supply shortages), as well as some relevant effects (e.g., banning social media and censorship).

\section{Regularization Experiment}
\label{appendix:reg}
In this appendix, we present an additional simulation evaluation to study overfitting issues and the effect of regularization in our extraction model. 

We addressed potential overfitting issues by adding a regularization term to the objective of the linear program. However, we have not shown the actual effects of overfitting when this term is not considered nor how this addition properly curtails overfitting issues. Preliminary qualitative evaluations showed that attempting to solve certain tasks led to overly complex maps after performing semantic interactions. Thus, we added a regularization term into the extraction model to minimize the risks of overfitting. In particular, the regularization term added to Figure \ref{fig:extraction_lp} acts like $L_1$ regularization. This term seeks to minimize the sum of all edge weights, leading to a sparse solution in the number of edges \cite{donoho2005sparse}. This provides two-fold benefits: first, it simplifies the resulting map, making it easier to understand, and second, it reduces the likelihood of overfitting (i.e., generating an overly complex map after using SI). 

Following these qualitative evaluations, we ran the simulation experiments again to compare both the basic model without regularization and the regularized model. For the purposes of this simulation-based evaluation, we use the simplified version of the Cuban protests data set with only 160 documents instead of the full data set with 500 documents (i.e., the same data used in the case study of Appendix \ref{appendix:case}), as running the full-scale simulation is a much more computationally expensive endeavor.

In this context, we first need to define a way to measure overfitting in the context of narrative map extraction. We do this by measuring the \textit{change in complexity} of the map after using SI. A drastically higher complexity value is associated with a higher cognitive load when analyzing the map (e.g., consider the difficulty of using a map with triple the number of edges compared to the baseline). To measure the complexity of the map we use the number of nodes and edges as our main metrics. We note that these complexity values do not have an inherent target or optimal value and their values depend on the map size parameter ($K$) used in the extraction. However, changes in these complexity measures after interacting with the narrative map are related to our intuitive understanding of changes in narrative complexity. Hence, their values must be evaluated relative to a baseline---the initial map before using semantic interactions.

\begin{figure*}[!htb]
    \centering
    \includegraphics[width=0.85\textwidth]{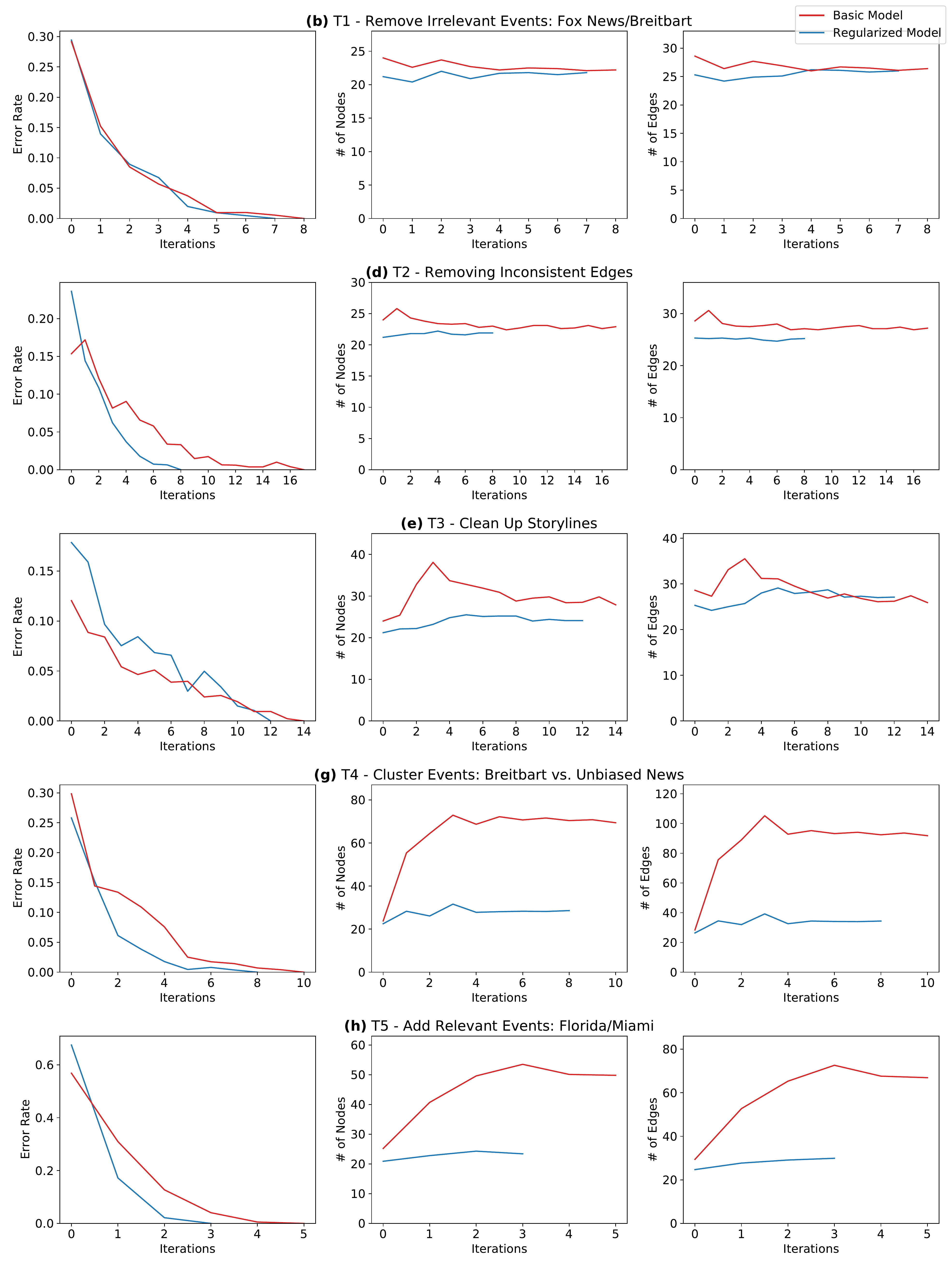}
    \caption{Overfitting analysis for the basic model vs. the regularized model. The left column shows the error rate plots. The middle column shows plots for the number of nodes throughout the simulation. The right column shows plots for the number of edges throughout the simulation. Both models converge towards zero error rate, but the basic model has overfitting issues in tasks T4 and T5 based on its drastic increases in graph complexity metrics (number of nodes and edges).}
    \label{fig:overfit}
\end{figure*}

We summarize our results in Figure \ref{fig:overfit}. Due to space constraints, we only show one example per task for those tasks that had multiple examples. The first column compares the error rate for the basic model and the regularized model. The second and third columns show the average number of nodes and edges, respectively, for each model. 

We note that the error rate converges to zero as expected in both models. There are no apparent differences in the rate of convergence between the two models. Thus, both models are able to capture user intent, just like in the full-scale simulation with the larger version of the data set. Regarding graph complexity, we note that, in general, the basic model usually has a slightly higher base complexity. This is expected, as regularization should reduce the model complexity in general. However, the difference is not particularly significant.

Next, we note that in tasks T1 and T2 both models behave similarly in terms of complexity---there are no significant changes in the number of edges or nodes throughout the simulation. However, we note that for T2 the regularized model converges significantly faster compared to the basic model. This faster rate of convergence could be due to the edge-based nature of task T2. Since the regularized model tries to construct sparser maps, the maps are less likely to contain inconsistent edges. 

Regarding T3, we note that the basic model has overfitting issues. The number of nodes increases compared to the initial map and remains higher throughout the simulation. The number of edges increases initially but then returns to lower levels. In contrast, the regularized model only shows a mild increase compared to the basic model. 

For T4 and T5, we note that the basic model has drastic increases in the number of nodes and edges even after a single iteration. The regularized model also has an increase in these metrics, but the changes are much milder. Thus, following our interpretation of complexity, the basic extraction model with no regularization has overfitting issues for the clustering-based tasks, while the regularized model is able to reduce the overfitting impacts of the interactions. 
 
In general, we note that semantic interactions that only work on the structure space level do not seem to lead to significant overfitting issues when used in simple tasks (T1 and T2) and only mild overfitting issues when used in more complex tasks (T3). However, semantic interactions that work on the projection space can lead to significant overfitting issues (T4 and T5). Nevertheless, the addition of regularization minimizes the impact of semantic interactions on narrative map complexity and solves most overfitting issues under our complexity-based interpretation for all tasks, without significantly increasing the convergence of the error rate.

\end{document}